\documentclass[fleqn,usenatbib]{mnras}

\usepackage{newtxtext,newtxmath}

\usepackage[T1]{fontenc}
\usepackage{ae,aecompl}

\usepackage{hyperref}
\usepackage{graphicx}	
\usepackage{blindtext}  
\usepackage{natbib}
\usepackage{rotating,times,graphicx,latexsym}
\usepackage{color}
\usepackage{longtable}
\usepackage{lscape}
\usepackage{lipsum} 
\usepackage{array}
\usepackage{flafter}

\usepackage{orcidlink}

\usepackage{soul} 
\usepackage{xargs} 

\newcommand{\hii}{H{\sc ii}~}
\graphicspath{ {images/} }

\title[HOYS YSOs]{A survey for variable young stars with small telescopes: VIII - Properties of 1687 Gaia selected members in 21 nearby clusters}

\author[Dirk Froebrich et al.]{Dirk Froebrich\orcidlink{0000-0003-4734-3345}$^{1}$$^{\thanks{E-mail: df@kent.ac.uk}}$,
Aleks Scholz$^{2}$,
Justyn Campbell-White\orcidlink{0000-0002-3913-3746}$^{3}$,
Siegfried Vanaverbeke\orcidlink{0000-0003-0231-2676}$^{4,5,6}$, 
\newauthor 
Carys Herbert$^{1}$, 
Jochen Eisl\"offel\orcidlink{0000-0001-6496-0252}$^{7}$, 
Thomas Urtly$^{8}$$^{\thanks{HOYS Observer}}$, 
Timothy P. Long$^{9}$$^{\dagger}$,
Ivan L. Walton$^{8}$$^{\dagger}$, 
\newauthor
Klaas Wiersema$^{10,11}$$^{\dagger}$,
Nick J. Quinn$^{8}$$^{\dagger}$, 
Tony Rodda$^{8}$$^{\dagger}$,  
Juan-Luis González-Carballo$^{12,13,14}$$^{\dagger}$,
\newauthor
Mario Morales Aimar$^{13,15}$$^{\dagger}$, 
Rafael Castillo García$^{13,14,16}$$^{\dagger}$,  
Francisco C. Sold\'{a}n Alfaro$^{13,14,17}$$^{\dagger}$,
\newauthor 
Faustino Garc\'{i}a de la Cuesta$^{18}$$^{\dagger}$,
Domenico Licchelli$^{19,20}$$^{\dagger}$, 
Alex Escartin Perez$^{13,21,22}$$^{\dagger}$, 
\newauthor 
Jos\'{e} Luis Salto Gonz\'{a}lez$^{13,23,24}$$^{\dagger}$,
Marc Deldem$^{14}$$^{\dagger}$,   
Stephen R.L. Futcher$^{8,14,25,26}$$^{\dagger}$,
\newauthor 
Tim Nelson$^{25,27}$$^{\dagger}$,
Shawn Dvorak$^{14,28}$$^{\dagger}$,
Dawid Mo\'{z}dzierski$^{29}$$^{\dagger}$,
Krzysztof Kotysz$^{29}$$^{\dagger}$,
\newauthor 
Przemys{\l}aw Miko{\l}ajczyk$^{29,30}$$^{\dagger}$, 
George Fleming$^{8}$$^{\dagger}$, 
Mark Phillips$^{8,31}$$^{\dagger}$, 
Tony Vale$^{8,14,32,33,34}$$^{\dagger}$,
\newauthor 
Yenal \"{O}{\u g}men$^{35}$$^{\dagger}$, 
Franky Dubois$^{4,5}$$^{\dagger}$, 
Samantha M. Rolfe$^{36}$$^{\dagger}$,
David A. Campbell$^{36}$$^{\dagger}$, 
\newauthor 
Heinz-Bernd Eggenstein$^{37}$$^{\dagger}$,
Franz-Josef Hambsch$^{5,14,38,39}$$^{\dagger}$, 
Michael A. Heald$^{14}$$^{\dagger}$,  
\newauthor 
Pablo Lewin\orcidlink{0000-0003-0828-6368}$^{40}$$^{\dagger}$,
Adam C. Rose$^{14}$$^{\dagger}$, 
Geoffrey Stone\orcidlink{0000-0001-5888-9162}$^{14,41}$$^{\dagger}$,
Martin Valentine Crow$^{8,42,43}$$^{\dagger}$,
\newauthor 
Simon Francis Dawes$^{8,43}$$^{\dagger}$,
Derek OKeeffe$^{44}$$^{\dagger}$,
Adam Popowicz$^{45}$$^{\dagger}$, 
Krzysztof Bernacki$^{45}$$^{\dagger}$,
\newauthor 
Andrzej Malcher$^{45}$$^{\dagger}$, 
Slawomir Lasota$^{45}$$^{\dagger}$,  
Jerzy Fiolka$^{45}$$^{\dagger}$, 
Adam Dustor$^{46}$$^{\dagger}$,
\newauthor 
Amritanshu Vajpayee\orcidlink{0000-0003-0718-1708}$^{47,48}$$^{\dagger}$,
Pat Devine$^{31}$$^{\dagger}$, 
Matthias Kolb$^{38,49}$$^{\dagger}$,
Jean-Baptiste Marquette$^{50}$$^{\dagger}$,
\newauthor 
Gregg L. Ruppel$^{51}$$^{\dagger}$,
Dan R. Crowson$^{52}$$^{\dagger}$,
Cledison Marcos da Silva$^{14,53,54}$$^{\thanks{HOYS Data Processor}}$, 
\newauthor 
Michel Michaud$^{55}$$^{\ddagger}$, 
Aashini L. Patel$^{1}$$^{\thanks{Beacon Observer}}$, 
Matthew D. Dickers\orcidlink{0000-0001-9615-9101}$^{1}$$^{\mathsection}$,
Lord Dover$^{1}$$^{\mathsection}$,
\newauthor 
Ivana I. Grozdanova$^{1}$$^{\mathsection}$,
James S. Urquhart\orcidlink{0000-0002-1605-8050}$^{1}$,
Chris J.R. Lynch$^{1}$,
\\
$^{1}$Centre for Astrophysics and Planetary Science, School of Physics and Astronomy, University of Kent, Canterbury CT2 7NH, UK\\
$^{2}$SUPA, School of Physics \& Astronomy, University of St Andrews, St Andrews KY16 9SS, UK\\
$^{3}$European Southern Observatory, Karl-Schwarzschild-Strasse 2, 85748 Garching bei M\"unchen, Germany\\
$^{4}$Public observatory ASTROLAB IRIS, Provinciaal Domein “De Palingbeek”, Verbrandemolenstraat 5, 8902 Zillebeke, Ieper, Belgium\\
$^{5}$Vereniging Voor Sterrenkunde (VVS), Oostmeers 122 C, 8000 Brugge, Belgium\\
$^{6}$Centre for Mathematical Plasma-Astrophysics, Department of Mathematics, KU Leuven, Celestijnenlaan 200B, 3001 Heverlee, Belgium\\
$^{7}$Th\"uringer Landessternwarte, Sternwarte 5, D-07778 Tautenburg, Germany\\
$^{8}$British Astronomical Association, Variable Star Section, PO Box 702, Tonbridge, TN9 9TX, UK\\ 
$^{9}$Timtek Systems Limited, 16 Laxton Way, Canterbury, Kent CT1 1FT, UK\\
$^{10}$Centre for Astrophysics Research, University of Hertfordshire, Hatfield, AL10 9AB, UK\\
$^{11}$School of Physics and Astronomy, University of Leicester, University Road, Leicester LE1 7RH, UK\\
$^{12}$Cerro del Viento Observatory (MPC~I84), Fernandez Pirfano Square 3, 06010 Badajoz, Spain\\
$^{13}$Observadores de Supernovas (ObSN)$^{\thanks{\tt \href{https://www.obsn.es/}{Observadores de Supernovas, https://www.obsn.es/}}}$, Spain\\
$^{14}$American Association of Variable Star Observers (AAVSO), 185 Alewife Brook Parkway, Suite 410, Cambridge, MA 02138, USA\\ 
$^{15}$Observatorio de Sencelles, Sonfred Road 1, 07140 Sencelles, Mallorca, Spain\\
$^{16}$Asociacion Astronomica Cruz del Norte, Calle Caceres 18, 28100 Alcobendas, Madrid, Spain\\
$^{17}$Science Department, Seville University, Av. de la Ciudad Jard\'{i}n, 20-22, 41005 Sevilla, Spain\\
$^{18}$La Vara, Valdes Observatory( MPC~J38), Barrio La Bara, sin n\'{u}mero Mu\~{n}as de Arriba c\'{o}digo postal 33784, Asturias, Spain\\
$^{19}$R.P. Feynman Observatory, Piazzetta del Ges\'{u} 3, 73034, Gagliano del Capo, Italy\\
$^{20}$Center for Backyard Astrophysics (CBA), Piazzetta del Ges\'{u} 3, 73034, Gagliano del Capo,  Italy\\
$^{21}$Belako, Aritz Bidea No 8, 4B Mungia Bizkaia, Basque Country, Spain\\
$^{22}$Hosting Trevinca, Agrupaci\'{o}n Astron\'{o}mica Vizcaina, Bizkaiko Astronomia Elkartea AAVBAE.net, c/ Iparragirre 46, 4º - dpto 2, Bilbo, Basque Country, Spain\\
$^{23}$Sociedad Malague\~{n}a de Astronom\'{i}a (SMA), Centro Cultural Jos\'{e} Mar\'{i}a Guti\'{e}rrez Romero, Cl Rep\'{u}blica Argentina, no 9, Urb. El Limonar, 29016 M\'{a}laga, Spain\\
$^{24}$Cal Maciarol Observatory, Cam\'{i} de l'Observatori S/N, 25691 \`{A}ger, Spain\\
$^{25}$Hampshire Astronomical Group, Hinton Manor Lane, Clanfield, PO8 0QR, UK\\
$^{26}$Royal Astronomical Society, Burlington House, Piccadilly, London W1J 0BQ, UK\\
$^{27}$Horndean Observatory, 6 Falcon Road, Horndean, Waterlooville, Hampshire, PO89BY, UK\\
$^{28}$Rolling Hills Observatory, Clermont, FL 34711, USA\\
$^{29}$Astronomical Institute, University of Wroc{\l}aw, ul. M. Kopernika 11, 51-622 Wroc{\l}aw, Poland\\
$^{30}$Astronomical Observatory, University of Warsaw, Al. Ujazdowskie 4, 00-478 Warsaw, Poland\\
$^{31}$Astronomical Society of Edinburgh, Edinburgh, UK\\
$^{32}$Wiltshire Astronomical Society, The Knoll, Lowden Hill, Chippenham SN15 2BT, UK\\
$^{33}$Bath Astronomers, 19 New King Street, Bath BA1 2BL, UK\\
$^{34}$The Herschel Society, The Herschel Museum of Astronomy, 19 New King Street, Bath BA1 2BL, UK\\
$^{35}$Green Island Observatory, Karao{\u g}laono{\u g}lu Street 63A, Ge\c{c}itkale Ma{\u g}usa, North Cyprus\\ 
$^{36}$Bayfordbury Observatory, School of Physics, Engineering and Computer Science, University of Hertfordshire, Lower Hatfield Road, Bayfordbury, Hertfordshire SG13 8LD, UK\\
$^{37}$Volkssternwarte Paderborn e.V., Im Schlosspark 13, 33104 Paderborn, Germany \\
$^{38}$Bundesdeutsche Arbeitsgemeinschaft für Ver\"{a}nderliche Sterne (BAV), Munsterdamm 90, 12169 Berlin, Germany\\
$^{39}$Groupe Europ\'{e}en d’Observations Stellaires (GEOS), 23 Parc de Levesville, 28300 Bailleau l’Ev\^{e}que, France\\
$^{40}$The Maury Lewin Astronomical Observatory, 420 N. Grand Ave., Glendora CA 91741, USA\\
$^{41}$First Light Observatory Systems, 19807 NE 391st St, Amboy, WA 98601, USA\\
$^{42}$Burnham Observatory, 19 Alexandra Road, Burnham on Crouch, Essex CM0 8BW, UK\\
$^{43}$Crayford Manor House Astronomical Society, Parsonage Lane Pavilion, Parsonage Lane, Sutton-at-Hone, Dartford DA4 9HD, Kent, UK\\
$^{44}$Cork Astronomy Club, 15 Ashdale Park, S Douglas rd. Cork, Ireland\\
$^{45}$Department of Electronics, Electrical Engineering and Microelectronics, Silesian University of Technology, Akademicka 16, 44-100 Gliwice, Poland\\
$^{46}$Department of Telecommunications and Teleinformatics, Silesian University of Technology, Akademicka 16, 44-100 Gliwice, Poland\\
$^{47}$Ignited Minds VIPNET Club (VP-UP0103), 1/125-G, New Civil Lines, Fatehgarh, Farrukhabad, Uttar Pradesh, India - 209 601\\
$^{48}$Akash Ganga: Centre for Astronomy, A/2, East \& West Villa, Nowroji Vakil Street, Grant Road, Mumbai, Maharashtra, India - 400 007\\
$^{49}$Sternwarte Neanderhoehe Hochdahl, Sedentaler Str. 105, 40699 Erkrath, Germany\\
$^{50}$Soci/'{e}t/'{e} Astronomique de France, 3 rue Beethoven, 75016 Paris, France\\
$^{51}$Dark Skies New Mexico Observatory, 1984 W. Golden Rose Pl, Oro Valley, AZ 85737, USA\\
$^{52}$Astronomical Society of Eastern Missouri (ASEM), 5 Nightfall Court, Dardenne Prairie MO 63368, USA\\
$^{53}$Uni\~{a}o Brasileira de Astronomia (UBA),  Rua 227, N° 25, Setor Leste Universit\'{a}rio, Goi\^{a}nia, Goi\'{a}s, CEP 74605-080, Brasil\\
$^{54}$Variable Stars South (VSS), Royal Astronomical Society of New Zealand, PO Box 3181, Wellington, New Zealand\\
$^{55}$MCD Observatory, 23 Langlois, Saint-Anaclet, Quebec, Canada, G0K1H0\\
}
\date{Accepted XXX. Received YYY; in original form ZZZ}

\pubyear{2023}

\begin{document}
\label{firstpage}
\pagerange{\pageref{firstpage}--\pageref{lastpage}}
\maketitle

\begin{abstract}

The Hunting Outbursting Young Stars (HOYS) project performs long-term, optical, multi-filter, high cadence monitoring of 25 nearby young clusters and star forming regions. Utilising Gaia~DR3 data we have identified about 17000 potential young stellar members in 45 coherent astrometric groups in these fields. Twenty one of them are clear young groups or clusters of stars within one kiloparsec and they contain 9143 Gaia selected potential members. The cluster distances, proper motions and membership numbers are determined. We analyse long term ($\approx$ 7~yr) V, R, and I-band light curves from HOYS for 1687 of the potential cluster members. One quarter of the stars are variable in all three optical filters, and two thirds of these have light curves that are symmetric around the mean. Light curves affected by obscuration from circumstellar materials are more common than those affected by accretion bursts, by a factor of 2~--~4. The variability fraction in the clusters ranges from 10 to almost 100 percent, and correlates positively with the fraction of stars with detectable inner disks, indicating that a lot of variability is driven by the disk. About one in six variables shows detectable periodicity, mostly caused by magnetic spots. Two thirds of the periodic variables with disk excess emission are slow rotators, and amongst the stars without disk excess two thirds are fast rotators -- in agreement with rotation being slowed down by the presence of a disk.

\end{abstract}

\begin{keywords}
stars: formation, pre-main sequence -- stars: variables: T\,Tauri, Herbig Ae/Be -- stars: rotation
\end{keywords}




\section{Introduction}

Variability is a characteristic feature of young stellar objects \citep[YSOs,][]{1945ApJ...102..168J}. It is caused by strong magnetic activity (i.e. spots and flares), accretion from protoplanetary disks and its variations, or obscurations by disk material. In many objects multiple sources of variability can be identified. Because variability is prevalent in YSOs, it is a helpful indicator to identify regions of active star formation, even with only a few epochs of data. Beyond that, extensive monitoring campaigns have been used to study rotation, accretion, and disk structure in YSOs \citep[see Protostars and Planets reviews by:][]{2007prpl.conf..297H, 2007prpl.conf..479B, 2014prpl.conf..433B, 2014prpl.conf..387A, 2023ASPC..534..355F}

Variability studies of young stars are ideally conducted quasi-simultaneously in multiple bands in the optical, to gain information on colour changes, which helps to disentangle the various sources of photometric variations. The CSI~2264 program \citep[e.g.][]{2014AJ....147...82C} has set the tone for comprehensive multi-filter campaigns, including optical and infrared photometry, albeit limited to one region and a 30~d time window. To obtain a comprehensive picture, the typical timescales of the variability, from hours to years, need to be covered. In the past, datasets that combine these two characteristics were rarely available. In particular, the long-term coverage was often missing, or only available in one band \citep[with exceptions, see ][]{2007A&A...461..183G}. The optical time-domain coverage of the sky has much improved over the last decade, thanks to dedicated missions to find supernovae \citep[e.g. ASAS-SN:][]{2017PASP..129j4502K} or to hunt for exoplanets (e.g. Kepler, TESS), and thanks to the time-domain capacity of Gaia. None of these however is particularly well suited for YSO variability. Therefore, there is still a need for dedicated multi-filter, multi-timescale studies of specific star forming regions.

The project Hunting Outbursting Young Stars (HOYS) is one of the endeavours designed to overcome this deficit, by collecting observations from a multitude of small telescopes from professional and amateur observatories, all using a pre-defined observing strategy and a specific set of target regions \citep[see overview paper by ][]{2018MNRAS.478.5091F}. In the first HOYS papers we have, for example, presented a study of rotation in one star forming region \citep{2021MNRAS.506.5989F}, a follow-up study of spot properties in the same region \citep{2023MNRAS.520.5433H}, and an analysis of the variability in a specific source that is caused by eclipses from structures in a protoplanetary disk \citep{2020MNRAS.493..184E}, among other findings. In this paper we take a more holistic view -- we aim to select prospective young stars in all 25 HOYS target regions, using data from Gaia, collate information about these samples, and study general variability properties among those samples. The precise kinematic information from Gaia is key for this paper, as it enables a systematic and robust selection of members of young clusters based primarily on kinematic information, parallax and proper motion, as demonstrated for example by \citet{2020ApJ...899..128K, 2021MNRAS.503.3232P, 2022AJ....164...57K,2023AJ....165..269L}.

This paper is structured as follows. In Sect.\,\ref{dataanalysis} we detail the analysis methods used to identify cluster members, obtain their light curves and characterise them. We discuss the light curve properties in general and on a cluster by cluster basis in Sect.\,\ref{results}.

\section{Data Analysis}\label{dataanalysis}

In this section we describe the selection of our samples using kinematics and photometry, and the light curve analysis. In all steps, we aim to identify reliable cluster members with good photometry in the HOYS database, by including several conservative limits. That means in turn that our samples do not constitute a complete list of cluster members for individual regions -- this is not the goal of this analysis.

\subsection{Cluster selection}

We are basing our selection of cluster members on the astrometry supplied by Gaia~DR3 \citep{2016A&A...595A...1G,2023A&A...674A...1G}. For each of the 25 HOYS fields\footnote{HOYS Target list: \url{https://hoys.space/target-lists/}}, we downloaded all Gaia sources within a radius of 0.6\,deg from the cluster centre. All sources with a parallax of less than 0.3\,mas, a parallax signal to noise ratio below five and Gmag fainter than 18\,mag were removed from our analysis. This ensures only stars with accurate distances and photometry are included in the analysis. Furthermore, we remove all blue sources with colours BP~$-$~Gmag~$<$~$-$0.2~mag and Gmag~$-$~RP~$<$~0.0~mag. This is the colour range for white dwarfs and cataclysmic variables, which we do not expect in our target regions. 

\begin{figure*}
\centering
\includegraphics[angle=0,height=7.0cm]{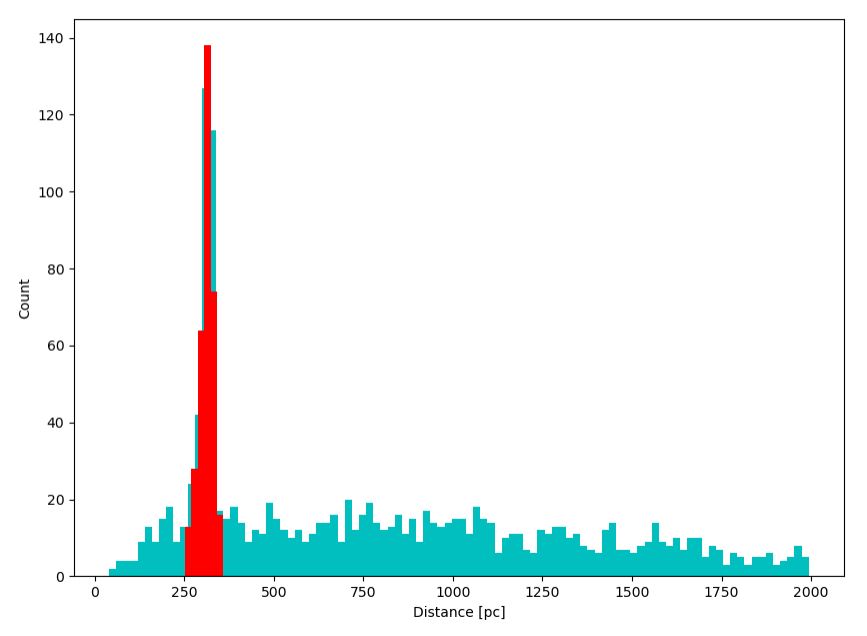} \hfill
\includegraphics[angle=0,height=7.0cm]{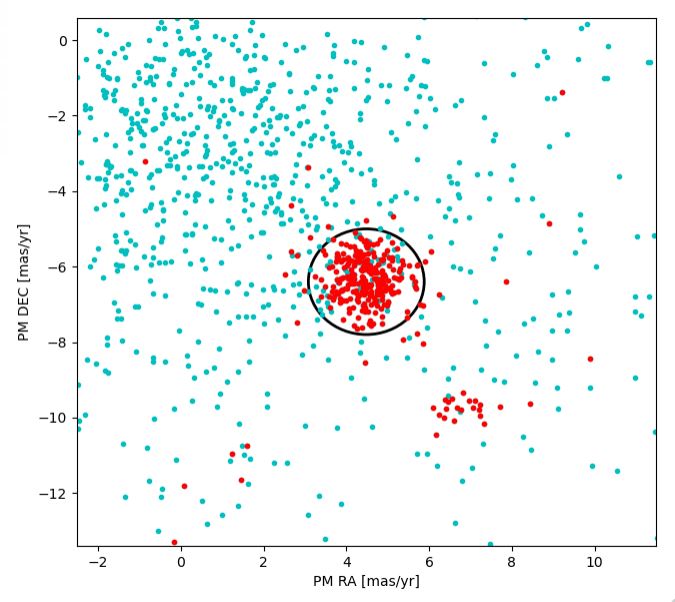} \\
\caption{ Example of the cluster member selection in Gaia~DR3 data for the main cluster in the IC~348 field. The {\bf left} panel shows the distance histogram with the stars selected based on their parallax highlighted in red. In the {\bf right} panel the proper motion values for all stars in the field are shown, with the sources selected in the left panel highlighted in red. The black circle encompasses the proper motion range used in the initial manual selection. The red-highlighted group of stars to the bottom right from the main group represents a secondary population of stars in this field - see Table\,\ref{gaiatable} for details on its properties. \label{cl_sel}}
\end{figure*}

We manually identified the peaks in the histogram of the distances (assumed to be one over the parallax) for each field, using a bin width of 20\,pc. All stars around each peak were selected. The distance range selected for each cluster is listed in Table~\ref{gaiatable}. The proper motion distributions for the selected stars were searched for groups with coherent proper motions. All stars within a group showing coherent parallax and proper motion were selected as potential cluster members. Figure\,\ref{cl_sel} shows one example (the main cluster in the IC~348 field) for this manual selection process. In Table\,\ref{gaiatable} we list for each field and cluster the ranges for distances and proper motion used in the selection. For each group we determined the median parallax and proper motion, together with their standard deviation and the standard error of the median. These are also listed in Table\,\ref{gaiatable}. Note that for some clusters we only used a sub-population within a small region centred around the cluster coordinates to estimate the median and scatter for distance and proper motion. That radius is listed in Table\,\ref{gaiatable}. The final selection of potential Gaia detected cluster members, i.e. all stars that are within three sigma from the median values in distance and proper motion, was done within 0.6\,deg from the cluster centre. The number of these Gaia selected potential cluster members is also listed in Table\,\ref{gaiatable}.

\begin{table*}
\caption{\label{gaiatable} This table contains the results from the Gaia~DR3 data selection for all HOYS fields. The first column is the cluster name. The main cluster (a) is the most populous young cluster associated with the HOYS target field. All other sub-populations are numbered consecutively and do have either a different proper motion and/or distance from the main cluster. In some cases the secondary clusters are known and we used their common names (e.g. NGC~7142 in the NGC~7129 field, or IC~4996 in the P~Cyg field). We further list: r: The radius around the central coordinates from which Gaia~DR3 sources are analysed when determining the cluster distance and proper motion; d$_1$, d$_2$: distance range in which cluster members are selected from; PM$_{\alpha}$, PM$_{\delta}$, PM$_{r}$: proper motions and range in which cluster members are selected from for distance and proper motion determination; $d$, $d^e$, $d^s$: median distance, standard error of the median, rms scatter of the cluster members; $\mu_{\alpha / \delta}$, $\mu_{\alpha / \delta}^e$, $\mu_{\alpha / \delta}^s$: median proper motion in RA/DEC, standard error of the median, rms scatter of the cluster members; N Gaia man.: number of Gaia stars selected manually for the cluster property determination; N Gaia 3~$\sigma$: number of Gaia stars selected as potential cluster members; N LC: number of stars with HOYS light curves; incl? Y/N: inclusion flag; Does the colour magnitude diagram look like a young cluster Y/N? / Is the cluster closer than 1~kpc Y/N? / Are there more than 20 stars with HOYS light curves Y/N?}
\centering
\setlength{\tabcolsep}{2.7pt}
\renewcommand{\arraystretch}{0.9}
\begin{tabular} {|c|c|cc|ccc|ccc|ccc|ccc|cc|c|c|}
\hline
Name & r & d$_1$ & d$_2$ & PM$_{\alpha}$ & PM$_{\delta}$ & PM$_{r}$ & $d$ & $d^e$ & $d^s$ & $\mu_{\alpha}$ & $\mu_{\alpha}^e$ & $\mu_{\alpha}^s$ & $\mu_{\delta}$ & $\mu_{\delta}^e$ & $\mu_{\delta}^s$ & \multicolumn{2}{c|}{N Gaia} & N & incl? \\ 
 & [deg] & [pc] & [pc] & \multicolumn{3}{c|}{[mas/yr]} & [pc] & [pc] & [pc] & \multicolumn{3}{c|}{[mas/yr]} & \multicolumn{3}{c|}{[mas/yr]} & man. & 3~$\sigma$ & LC & Y/N \\  \hline
\multicolumn{20}{|l|}{Clusters/sub-groups used in analysis} \\ \hline
IC~348~a & 0.3 & 250 & 360 & 4.5 & -6.4 & 1.4 & 318.1 & 1.19 & 23.6 & 4.478 & 0.039 & 0.524 & -6.345 & 0.039 & 0.518 & 178 & 272 & 70 & YYY \\ 
IC~348~b & 0.6 & 250 & 360 & 6.8 & -9.4 & 1.0 & 297.0 & 4.0 & 17.9 & 6.723 & 0.097 & 0.433 & -9.738 & 0.057 & 0.253 & 20 & 33 & 5 & YYN \\
$\lambda$-Ori~a & 0.4 & 330 & 450 & 0.7 & -2.0 & 0.4 & 398.9 & 1.6 & 14.8 & 0.780 & 0.018 & 0.168 & -2.005 & 0.017 & 0.154 & 85 & 191 & 25 & YYY \\
$\lambda$-Ori~b & 0.6 & 330 & 450 & 1.8 & -2.1 & 0.4 & 403.7 & 1.85 & 13.5 & 1.652 & 0.024 & 0.172 & -2.138 & 0.021 & 0.151 & 53 & 91 & 7 & YYN \\
M~42~a & 0.3 & 300 & 500 & 1.3 & -0.1 & 3.0 & 400.4 & 0.89 & 23.2 & 1.307 & 0.029 & 0.76 & 0.215 & 0.04 & 1.036 & 684 & 2157 & 250 & YYY \\
L~1641~N~a & 0.3 & 320 & 440 & 1.0 & -0.3 & 1.4 & 388.4 & 1.29 & 16.8 & 1.034 & 0.038 & 0.49 & 0.327 & 0.045 & 0.579 & 168 & 978 & 97 & YYY \\
$\sigma$-Ori~a & 0.3 & 350 & 450 & 1.7 & -0.4 & 0.7 & 404.5 & 1.36 & 15.0 & 1.618 & 0.031 & 0.347 & -0.491 & 0.026 & 0.29 & 122 & 210 & 61 & YYY \\
$\sigma$-Ori~c & 0.6 & 330 & 400 & 1.9 & -1.6 & 0.5 & 368.5 & 1.68 & 12.9 & 1.803 & 0.022 & 0.165 & -1.475 & 0.023 & 0.175 & 59 & 132 & 16 & YYN \\
NGC~2264~a & 0.6 & 650 & 850 & -1.5 & -3.7 & 0.5 & 732.0 & 2.1 & 38.6 & -1.597 & 0.01 & 0.181 & -3.635 & 0.011 & 0.201 & 339 & 423 & 163 & YYY \\
NGC~2264~b & 0.6 & 650 & 850 & -2.5 & -3.7 & 0.5 & 737.3 & 2.74 & 41.5 & -2.425 & 0.015 & 0.231 & -3.682 & 0.012 & 0.187 & 229 & 302 & 128 & YYY \\
V~898~Ori~a & 0.6 & 300 & 500 & 0.5 & -0.3 & 1.4 & 402.8 & 3.77 & 35.0 & 0.222 & 0.065 & 0.601 & -0.31 & 0.064 & 0.592 & 86 & 194 & 22 & YYY \\
YY~Ori~a & 0.3 & 330 & 430 & 1.3 & 0.5 & 0.5 & 386.9 & 0.94 & 13.4 & 1.29 & 0.013 & 0.179 & 0.553 & 0.014 & 0.202 & 204 & 712 & 66 & YYY \\
V~555~Ori~a & 0.3 & 340 & 460 & 1.2 & 0.3 & 1.9 & 393.4 & 0.88 & 16.8 & 1.229 & 0.028 & 0.543 & -0.275 & 0.043 & 0.833 & 367 & 1569 & 118 & YYY \\
IC~5070~a & 0.6 & 750 & 900 & -1.4 & -3.1 & 0.75 & 832.4 & 2.7 & 33.6 & -1.328 & 0.028 & 0.347 & -3.076 & 0.026 & 0.320 & 155 & 252 & 84 & YYY \\
IC~5070~b & 0.6 & 750 & 900 & -0.9 & -4.3 & 0.7 & 824.7 & 4.43 & 36.0 & -0.977 & 0.036 & 0.294 & -4.148 & 0.041 & 0.335 & 66 & 114 & 47 & YYY \\
IC~1396~A~a & 0.6 & 850 & 1050 & -2.4 & -4.8 & 0.8 & 942.0 & 2.39 & 45.4 & -2.418 & 0.016 & 0.311 & -4.724 & 0.015 & 0.293 & 361 & 549 & 221 & YYY \\
IC~1396~N~a & 0.6 & 850 & 1050 & -1.9 & -4.1 & 1.2 & 955.0 & 3.1 & 43.9 & -2.021 & 0.03 & 0.428 & -3.964 & 0.04 & 0.568 & 200 & 642 & 168 & YYY \\
IC~1396~N~b & 0.6 & 580 & 650 & -6.2 & -5.9 & 0.4 & 619.8 & 2.69 & 10.8 & -6.139 & 0.031 & 0.124 & -5.826 & 0.031 & 0.125 & 16 & 30 & 13 & YYN \\
NGC~7129~a & 0.25 & 850 & 950 & -1.7 & -3.4 & 0.7 & 910.0 & 4.11 & 21.3 & -1.727 & 0.037 & 0.191 & -3.38 & 0.054 & 0.282 & 27 & 55 & 28 & YYY \\
IC~5146~a & 0.3 & 600 & 900 & -2.8 & -2.6 & 1.0 & 784.8 & 5.07 & 51.2 & -2.845 & 0.034 & 0.342 & -2.704 & 0.039 & 0.397 & 102 & 178 & 75 & YYY \\
\hline \multicolumn{20}{|l|}{Clusters/sub-groups identified but not analysed in detail} \\ \hline
$\sigma$-Ori~b & 0.6 & 350 & 450 & -2.2 & 1.1 & 0.9 & 421.9 & 3.66 & 19.0 & -2.135 & 0.063 & 0.328 & 1.27 & 0.066 & 0.341 & 27 & 49 & 15 & NYN \\
NGC~2068~a & 0.6 & 300 & 480 & -1.2 & -1.0 & 0.9 & 420.0 & 2.34 & 18.1 & -0.982 & 0.05 & 0.385 & -1.015 & 0.042 & 0.324 & 60 & 73 & 0 & NYN \\
NGC~2068~b & 0.6 & 300 & 480 & 0.1 & -0.6 & 0.8 & 417.9 & 2.44 & 17.8 & 0.066 & 0.051 & 0.369 & -0.555 & 0.044 & 0.323 & 53 & 69 & 0 & NYN \\
NGC~2068~c & 0.6 & 300 & 480 & -2.9 & 1.2 & 0.7 & 379.3 & 6.86 & 24.7 & -2.882 & 0.048 & 0.173 & 1.194 & 0.074 & 0.268 & 13 & 32 & 0 & NYN \\
NGC~2244~a & 0.3 & 1420 & 1600 & -1.7 & 0.2 & 0.5 & 1512.7 & 3.94 & 51.1 & -1.751 & 0.013 & 0.166 & 0.221 & 0.015 & 0.198 & 168 & 502 & 136 & YNY \\
NGC~2244~b & 0.6 & 590 & 710 & -1.7 & -4.7 & 0.6 & 658.3 & 4.52 & 26.0 & -1.707 & 0.033 & 0.188 & -4.762 & 0.027 & 0.154 & 33 & 69 & 18 & NYN \\
ASASSN-13DB~a & 0.6 & 290 & 440 & 1.3 & -1.0 & 0.5 & 387.3 & 3.16 & 18.4 & 1.306 & 0.029 & 0.171 & -0.964 & 0.042 & 0.244 & 34 & 57 & 0 & NYN \\
ASASSN-13DB~b & 0.6 & 290 & 440 & 0.5 & -1.4 & 0.6 & 346.2 & 5.28 & 23.6 & 0.367 & 0.043 & 0.193 & -1.24 & 0.057 & 0.254 & 20 & 37 & 0 & NYN \\
V~555~Ori~b & 0.6 & 340 & 460 & -1.8 & 1.4 & 0.7 & 419.4 & 3.75 & 14.5 & -1.737 & 0.047 & 0.182 & 1.401 & 0.056 & 0.215 & 15 & 25 & 0 & NYN \\
Gaia~17~bpi~a & 0.3 & 1200 & 1350 & -1.1 & -5.4 & 1.1 & 1300.3 & 6.26 & 42.9 & -0.997 & 0.049 & 0.338 & -5.654 & 0.058 & 0.396 & 47 & 289 & 39 & YNY \\
Gaia~19~fct~a & 0.3 & 1000 & 1300 & 0.2 & -0.1 & 0.35 & 1127.1 & 4.43 & 58.0 & 0.251 & 0.009 & 0.114 & -0.158 & 0.009 & 0.112 & 172 & 247 & 100 & YNY \\
Gaia~19~fct~b & 0.6 & 1000 & 1400 & -3.5 & 0.9 & 0.9 & 1230.9 & 7.61 & 88.1 & -3.565 & 0.037 & 0.424 & 1.017 & 0.028 & 0.319 & 134 & 411 & 57 & YNY \\
Gaia~19~eyy~a & 0.3 & 1300 & 1500 & -1.8 & 1.1 & 0.3 & 1405.5 & 5.48 & 45.2 & -1.771 & 0.011 & 0.092 & 1.107 & 0.011 & 0.089 & 68 & 131 & 50 & YNY \\
Gaia~19~eyy~b & 0.4 & 1200 & 1400 & -5.3 & 5.0 & 0.4 & 1274.2 & 5.69 & 39.0 & -5.407 & 0.016 & 0.109 & 4.96 & 0.023 & 0.155 & 47 & 101 & 21 & YNY \\
Gaia~19~eyy~c & 0.6 & 1900 & 3000 & -4.9 & 4.5 & 0.3 & 2487.9 & 25.09 & 257.1 & -4.872 & 0.01 & 0.1 & 4.476 & 0.01 & 0.103 & 105 & 274 & 22 & YNY \\
MWSC~3274~a & 0.6 & 830 & 970 & -1.1 & -1.6 & 0.3 & 884.5 & 3.45 & 29.9 & -1.145 & 0.012 & 0.101 & -1.589 & 0.009 & 0.081 & 75 & 93 & 30 & NYY \\
IC4996 & 0.25 & 1900 & 2400 & -2.6 & -5.3 & 0.3 & 2158.7 & 8.84 & 117.3 & -2.604 & 0.008 & 0.108 & -5.346 & 0.008 & 0.108 & 176 & 962 & 294 & YNY\\
Berkeley~86~a & 0.3 & 1600 & 2000 & -3.5 & -5.5 & 0.35 & 1839.0 & 7.82 & 92.2 & -3.451 & 0.012 & 0.137 & -5.479 & 0.013 & 0.149 & 139 & 990 & 217 & YNY \\
Berkeley~86~b & 0.4 & 1600 & 2000 & -3.2 & -6.0 & 0.2 & 1852.5 & 7.89 & 75.6 & -3.246 & 0.008 & 0.077 & -6.008 & 0.008 & 0.078 & 92 & 323 & 93 & YNY \\
Berkeley~86~c & 0.4 & 1440 & 1840 & -0.7 & -3.1 & 0.2 & 1636.0 & 12.66 & 86.8 & -0.695 & 0.01 & 0.069 & -3.081 & 0.009 & 0.061 & 47 & 113 & 16 & YNN \\
NGC~7142 & 0.30 & 2300 & 2800 & -2.7 & -1.4 & 0.25 & 2563.8 & 8.45 & 129.8 & -2.675 & 0.005 & 0.076 & -1.355 & 0.005 & 0.076 & 236 & 329 & 286 & NNY \\
IC~5146~b & 0.6 & 1100 & 1900 & 1.3 & -2.4 & 0.3 & 1531.0 & 15.61 & 120.9 & 1.276 & 0.017 & 0.135 & -2.376 & 0.014 & 0.106 & 60 & 126 & 40 & NNY \\
FSR~408~a & 0.6 & 750 & 950 & -0.8 & -2.3 & 1.0 & 860.2 & 1.78 & 41.9 & -0.876 & 0.017 & 0.405 & -2.29 & 0.014 & 0.325 & 556 & 2418 & 0 & NYN \\
FSR~408~b & 0.6 & 750 & 950 & -2.3 & -2.5 & 0.4 & 875.6 & 3.78 & 34.9 & -2.334 & 0.017 & 0.159 & -2.484 & 0.019 & 0.178 & 85 & 344 & 0 & NYN \\
\hline
\end{tabular}
\end{table*}

In many of the HOYS target regions this procedure resulted in more than one group of stars. In total 45 such groups or clusters of stars have been identified in the 25 HOYS target regions. These are all listed in Table\,\ref{gaiatable}. Typically there is a main cluster of objects dominating the field, which is the intended HOYS target. The secondary clusters are usually sub-populations with different proper motions at the same distance or fore/background clusters/groups in the same field of view. In Appendix\,\ref{app_clusters} we briefly discuss all the clusters/groups identified in each HOYS target field.

In total, about 17000 unique Gaia sources are selected as potential members of the 45 clusters. A sub-sample of these could in principle be considered as members of several of the clusters due to spatial and/or proper motion range overlap of the populations. Hence, every Gaia source has been assigned to the cluster to which it most likely belongs, based on the deviation of its parallax and proper motion values from the median value of all potential cluster members. For more details on the samples for individual clusters we refer again to Table\,\ref{gaiatable}. 

\subsection{HOYS light curve selection}

For each identified cluster/group we extracted all light curves from the HOYS database\footnote{HOYS database: \url{http://astro.kent.ac.uk/HOYS-CAPS/}}, on October 21st, 2022. Prior to this, all pairs of Gaia sources with separations of less than three arcseconds were identified and the fainter of the sources were removed. This was done to avoid false cross-matches of HOYS photometry to fainter Gaia targets, which are most likely not correct, given the typical seeing of three arcseconds in the HOYS data \citep{2018MNRAS.478.5091F,2020MNRAS.493..184E}. Only HOYS data points that had a photometric uncertainty below 0.2\,mag and seeing of better than 5\arcsec\ were used. Furthermore, only stars that had at least 100 photometry data points in each of the V, R, and I filters were considered for analysis. Furthermore, we find that light curves of stars that are within 5 arc-minutes from very bright sources (Gmag~$<$6~mag) can have unreliable photometry. These light curves have hence been removed from the analysis. All data were calibrated following the procedure described in \citet{2020MNRAS.493..184E}. This leaves just over 3000 V, R, and I-band light curves of potential cluster members for investigation. 

\begin{figure}
\centering
\includegraphics[angle=0,width=\columnwidth]{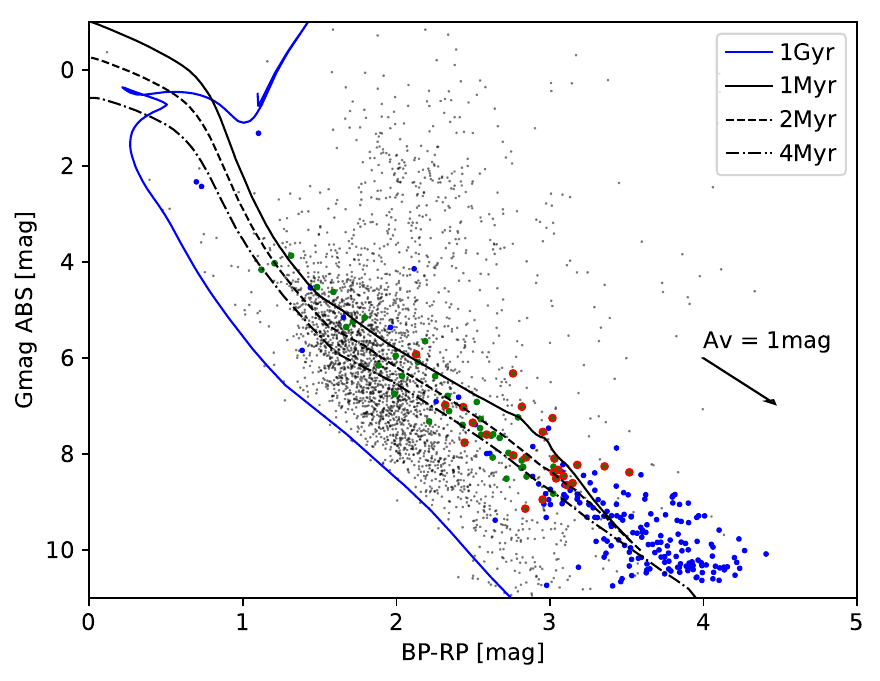} \hfill
\caption{ Example of a Gaia colour-magnitude diagram for the main cluster in the IC~348 field. We converted the apparent Gmag values into absolute magnitudes using each star's parallax value. Note that no extinction correction has been applied. All Gaia~DR3 sources in the field are shown as small gray dots, potential cluster members as large blue dots, all stars with HOYS light curves are shown as large green dots and all stars identified as variable are shown as large red dots. The over plotted isochrones are based on the PHOENIX models \citep{2013A&A...553A...6H}. The arrow represents an extinction of 1~mag in $A_V$ and is based on the average extinction coefficients for the Gaia filters from \citet{2018MNRAS.479L.102C}. \label{cl_cmd} }
\end{figure}

In Table\,\ref{gaiatable} we list for each cluster how many HOYS light curves are available for analysis. The Gaia colour-magnitude diagrams (CMDs) with overlaid isochrones \citep{2013A&A...553A...6H} were visually inspected for each cluster. Note that we do not apply an extinction correction for any of the CMDs. In Fig.\,\ref{cl_cmd} we show an example CMD for the main cluster in the IC\,348 field. Note that the extinction vector in these plots is almost parallel to the direction of the young (few Myr) isochrones near $BP-RP=2$~mag.  We selected all clusters that clearly showed a population of young stars, i.e. the members aligned with a young isochrone. Furthermore, all clusters with a median distance of the potential Gaia members of less than 1\,kpc were selected to ensure the samples of stars in all clusters are comparable in mass. All stars in these clusters were analysed in our general YSO sample. Finally, we selected all clusters with at least 20 HOYS light curves for a more detailed analysis on a cluster by cluster basis. All these selections are summarised in the final column in Table\,\ref{gaiatable}. There are spatial overlaps between some of the HOYS target fields. For the analysis on a cluster by cluster basis, we have hence merged these. In particular, the cluster M\,42 contains all stars from the HOYS fields M\,42, L\,1641\,N, V\,898\,Ori, YY\,Ori, and V\,555\,Ori; and IC\,1396 contains all objects from the IC\,1396\,A and IC\,1396\,N fields.

\subsection{HOYS light curve analysis}\label{lc_analysis}

In this section we briefly describe the parameters determined for each HOYS light curve for every selected potential cluster member. A visual inspection of the data showed that despite the strict quality selection applied to the photometry (see above), a few light curves contained a small number of erroneous, outlying data points. We define outliers as single brightness measurements that are either much brighter or much fainter than any of the other magnitudes of the star. In the absence of independent data taken in the same nights, we have to assume they are caused by photometric errors. Most of these occur in objects that are situated on a spatially highly variable background emission (as in e.g. M\,42). We remove all such points in all light curves if they were more than 3$\sigma$ away from the median magnitude in the light curve. Note that these could be correct measurements of short lived flares or very narrow dips. Furthermore, there is a small sub-set of light curves which do show highly unusual variability patterns which cannot be attributed to any specific issues in the photometry. These objects do remain in the sample, and might hence cause small, but statistically insignificant changes in the results due to potential mis-classification of their properties.

For each light curve and filter we determined a general Lomb-Scargle (GLS) periodogram \citep{1982ApJ...263..835S,2009A&A...496..577Z} using the implementation in {\tt astropy}. The periodograms use test frequencies between one over 1.3\,d and one over half the light curve length. They are distributed homogeneously in frequency space with a frequency spacing of one divided by 20 times the light curve length. Despite the many participants and their wide-spread geographic distribution \citep{2018MNRAS.478.5091F}, the observing cadence of the ground-based observatories involved in the HOYS project is usually close to 1\,d during periods of continuous clear weather. Therefore periodograms are often dominated by 1\,d periods and their harmonics/aliases. We therefore compute the window function of the light curves together with the periodograms \cite[following Eq.~A10 in][]{2021MNRAS.506.5989F}. We remove all periods with peaks of the window function which are higher than three standard deviations of the window function and their aliases from consideration in the periodograms.  After this step, the strongest period is logged for each object and photometric filter. The corresponding false-alarm probability (FAP) is computed using the bootstrapping implementation in {\tt astropy}. This is the same procedure as described in detail in \citet{2021MNRAS.506.5989F}. In a follow-up paper (Herbert et al., in prep.) we will analyse all periodic light curves in the entire sample of cluster members, including the identification of periods with a number of different period finding and evaluation methods, as detailed in \citet{2021MNRAS.506.5989F}.

In order to classify the light curves in the HOYS sample according to their general properties, we compute the variability metrics $Q$ and $M$ following \citet{2014AJ....147...82C}, which quantify the quasi-periodicity and flux asymmetry, respectively. 

\begin{equation} \label{eq_q}
Q=\frac{\sigma_{red}^{2}-\sigma_{phot}^{2}}{\sigma_{m}^{2}-\sigma_{phot}^{2}}
\end{equation}

The quasi-periodicity index $Q$ is defined in Eq.\,\ref{eq_q}. There, $\sigma_{red}^{2}$ is the variance of the data after a smoothed periodic component has been subtracted, $\sigma_{phot}$ is the measurement uncertainty (set to the average photometry uncertainty of all light curve data points) and $\sigma_{m}^{2}$ is the variance of the original light curve. The smoothed periodic component is determined using a boxcar filter with a width equal to one fourth of the period in a phase folded light curve. The $Q$ values close to zero indicate light curves which are close to strictly periodic, whereas $Q$ values close to 1 are found for signals that are completely stochastic. Values of the $Q$ index in between those extremes correspond to various levels of quasi-periodic behaviour.  

\begin{equation}  \label{eq_m}
 M=\frac{\langle m_{10\%} \rangle-m_{med}}{\sigma_{m}}  
\end{equation}

The flux asymmetry index $M$ is defined in Eq.\,\ref{eq_m}, where $\langle m_{10\%} \rangle$ denotes the mean of the 10 percent brightest and faintest data points in a given light curve, $m_{med}$ is the median magnitude and $\sigma_{m}$ is the standard deviation of the light curve.  Stars with negative values of $M$ are predominantly objects undergoing brightening, while stars with positive values of $M$ are predominantly stars going through dimming. Light curves with values of $M$ close to zero are variables with symmetric light curves.

\begin{figure}
\centering
\includegraphics[angle=0,width=\columnwidth]{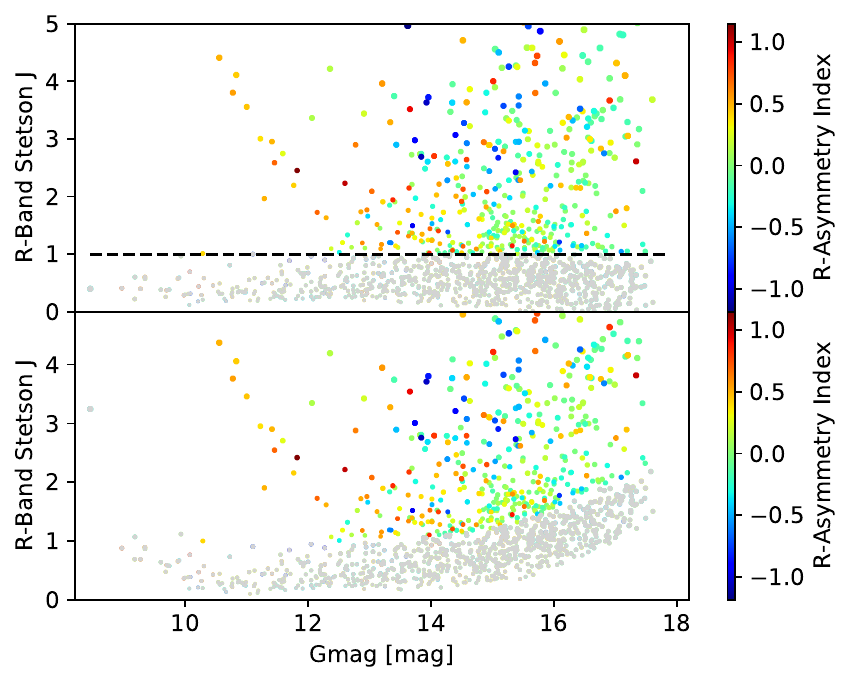} \\
\caption{ Stetson~J index distribution for the light curves of the investigated clusters, using the R-band data. In the {\bf bottom panel} the original values are displayed, which show a magnitude dependence. In the {\bf top panel} the values after our applied correction (see text for details) are shown and the dashed line marks Stetson~J~=~1, above which we consider stars variable. The colour code for all points represents the asymmetry index $M$, except for the non-variable objects which are shown in gray. \label{stet_cor}}
\end{figure}

For each light curve, we also compute the Stetson~J variability index \citep{1996PASP..108..851S} as a main indicator of the overall level of variability of a light curve. This index basically determines how much more variable compared to the photometric uncertainty a star is. As has been found in \citet{2020MNRAS.493..184E}, the HOYS calibration procedure slightly underestimates the photometric uncertainties for fainter stars. Thus, the Stetson~J index for our stars shows a slight magnitude dependence. We have thus fit a magnitude dependent running median to the Stetson~J index values and subtracted this function from the values. We then added the median of all Stetson~J indices in the magnitude range 12-15\,mag to all values to ensure the correct values for this range (which does not show any magnitude dependence) are restored. In Fig.\,\ref{stet_cor} we show the Stetson~J values for the R-Band data before and after the above described correction. Stars with values of this corrected Stetson~J index above one are considered variables in our sample. 

\subsection{Auxiliary Data and Definitions}\label{auxdata}

In order to investigate the evolutionary stages of the selected cluster members, we have utilised the cross matches to 2MASS \citep{2006AJ....131.1163S} and the WISE All-sky catalogue \citep{2013yCat.2328....0C} provided in Gaia~DR3, and downloaded the J, H, and K as well as the W1, W2, W3, and W4 magnitudes when available for all sources. Only data with uncertainties of less than 0.1~mag in WISE were included in the analyses.

In this work, we analyse the light curve properties of cluster members with and without a K-W2 excess. We consider sources with excess infrared emission, or detectable inner disk, as objects that have a K-W2 colour in excess of 0.5~mag \citep{2012A&A...540A..83T}. Objects with a bluer K-W2 colour will be referred to as non-excess sources or photospheres. 

Furthermore, we consider stars as variable if they have a Stetson~J index (corrected as detailed above) larger than one. All other objects are considered as non-variable. Light curves of variable stars with an $M$ index above 0.5 are considered as dippers. Objects with an $M$ index below $-$0.5 are classified as bursters. All other sources with an $M$ index between $-$0.5 and 0.5 are considered symmetric. Non-variable sources cannot be classified in this way. Note that in principle flare stars could be classified as bursters, if our data catches the objects during a number of such flare events. Furthermore, we use the above values to distinguish dippers, bursters, and symmetric light curves for consistency in following \citet{2014AJ....147...82C}. However, the distribution of $M$ values (as discussed in Sect.~\ref{res_varsym}) shows that there is a continuum of values with no clear 'under-population' at $\pm 0.5$. Thus, this is an arbitrary cut, and other works, such as the simulations by \citet{2021ApJ...908...16R} use different values ($\pm$0.25) for the cut and slightly different definitions for the $M$ index. Hence, absolute values for the fraction of dipper and burster light curves should be interpreted with caution.

Finally, variable objects with a $Q$ index between zero and 0.45 are classified as periodic objects. If the $Q$ index is between 0.45 and 0.85 the source is considered semi-periodic. Irregular or stochastic light curves have a $Q$ value between 0.85 and one. As for the $M$ index, non-variable sources cannot be classified based on the $Q$ index. The periods determined as part of the evaluation for the $Q$ index are considered significant if the associated FAP is below 1$\times 10^{-6}$. Following \citet{2021MNRAS.506.5989F} we use a period of 5.5~d to distinguish fast and slow rotators. The final note in the last paragraph, about the specific values distinguishing classes of sources also applies to the $Q$-values.

\section{Results}\label{results}

Our selection of potential clusters and groupings in the 25 HOYS target fields has identified 45 different clusters and/or groups of co-moving stars. In most fields these are dominated by a main cluster - the original target of the HOYS project. The median distances and proper motions for each of these groups are listed in Table\,\ref{gaiatable}. For the analysis in this paper, we only select clusters/groups which clearly appear young in the Gaia CMD (see Fig.\,\ref{cl_cmd} for an example) and have a distance of less than 1\,kpc. There are 21 such clusters and groups. All of these are listed in the top part of Table\,\ref{gaiatable}. All other groups are listed in the bottom part of the table for completeness. We will not analyse their potential members in this work.

In total, the 21 clusters/groups contain 9143 unique Gaia~DR3 sources as potential members. The HOYS database contains light curves with at least 100 photometry data points in each of the V, R, and I-Band filters for 1687 (18 percent) of these. This is what we consider the {\it full sample} of young stars. We are analysing the overall properties of this full sample and their light curves in Section~\ref{res_lcs}. In Section~\ref{res_cluster} we investigate the properties of the stars in the clusters that have at least 20 members. These clusters/groups are indicated with 'YYY' in the last column of Table\,\ref{gaiatable}.

\subsection{General light curve results}\label{res_lcs}

\subsubsection{Light curve variability and symmetry}\label{res_varsym}

We show the distribution of Stetson~J indices for the full sample of stars in the top panel of Fig.\,\ref{stet_cor} for the R-band data. The plots in the other filters are quantitatively similar. As one can see, the majority of sources is not variable. However, a fraction of 40/38/29 percent of the stars are considered variable with a Stetson~J index above one in V, R, and I, respectively. We consider these sources the {\it variable sample} of young stars in each filter. The drop in the I-band variability fraction is due to the fact that generally variability amplitudes for the majority of physical mechanisms (surface spots, line of sight extinction changes, mass accretion rate variations) are lower at longer wavelengths. Note that 25 percent of all stars are considered variable in all three filters simultaneously, and 44 percent are variable in at least one of the filters. In Fig.~\ref{stet_cor} we only show values for Stetson~J up to five, and a fraction of 3.7/5.0/3.0 percent of the stars is more variable than this, up to a maximum Stetson~J index of 25/34/56 in V, R, and I, respectively. The symbol colour in the plot indicates the asymmetry index $M$ for each star. Note that this is only meaningful for the variable sample, i.e. the sources with a Stetson~J index above one (see below). In Fig.~\ref{examplelc} in the Appendix we show a few examples of symmetric, dipper, and burster light curves.

The number of objects in our full and variable sample per magnitude bin starts dropping for Gmag~=~16~mag and fainter. Thus, the samples become incomplete for sources fainter than that limit. This is caused by the variations in limiting magnitude and field of view in our data. Similarly, objects brighter than Gmag~=~10~mag are at or near saturation in some of the data. Independent of this, or the filter used, the dominant type of light curve in the variable sample is symmetric ($-0.5 < M < +0.5$). Typically about 50 to 80 percent of the variable stars are categorised as symmetric (depending on Gmag and filter). Within the Gmag~=~10~--~14~mag range (where we are certainly complete), the fraction of symmetric light curves is 0.67/0.62/0.56 for V, R, and I, respectively. The remaining objects are bursters and dippers. The ratio of bursters to dippers for the variable sources in the Gmag~=~10~--~14~mag range is 0.26/0.62/0.00, for V, R, and I, respectively. Thus, in all filters, the brighter sources are dominated by dipper light curves, which outnumber the burster light curves by a factor of two to four or more. At fainter magnitudes (above about 14/15/15 in V, R, and I) the burster light curves start to dominate over dippers. 

All of these statistics are caused by the typical bias of a magnitude limited sample, like ours. Most of the young stars in our sample are red, i.e. they are brighter in the I-band than in the V-filter. Furthermore, there are many more fainter (less massive) stars than bright ones. Thus, many dipper stars will not be detected in their faint state in the V-band, and they hence get mis-classified as symmetric light curves. This is less of an issue in the R and I-band, since the objects are brighter in these. Similarly, some outbursting stars will be saturated in the I-band data but not in the V-filter, which can lead to an erroneous classification as symmetric light curve instead of burster in that filter. This issue is further influenced by the above discussed bias that changes in brightness due to accretion rate changes and/or line of sight extinction variations are smaller in the I-band. The R-band data hence suffers the least amount of bias. If we consider this, and only the magnitude range for which we most likely have a complete sample, then we find that dipper light curves amongst our variable sample outnumber the bursters by a factor of two to four. However, as indicated above, the symmetric light curves dominate the variable sample with 55~--~65 percent being classified as symmetric.

\begin{figure}
\centering
\includegraphics[angle=0,width=\columnwidth]{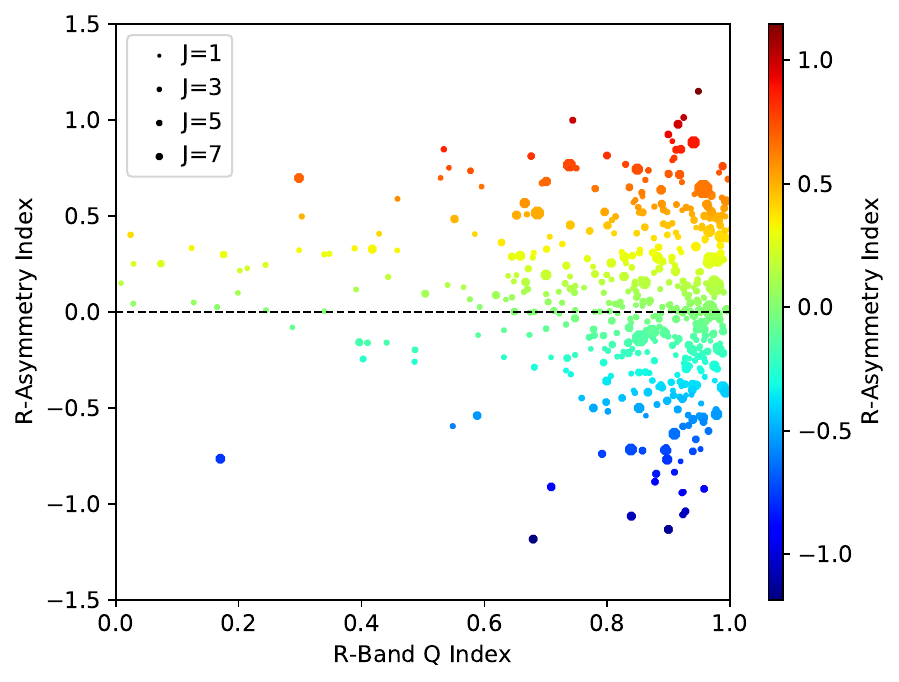} 
\caption{ R-band $Q$ index vs $M$ index for all variable sources. The colour code also represents the R-band asymmetry index $M$ and the symbol size is proportional to the R-band Stetson~J index for each source. The legend gives an indication of the respective symbol sizes. \label{qm-plot}}
\end{figure}

In Fig.\,\ref{qm-plot} we show the quasi-periodicity index $Q$ vs the asymmetry index $M$ of the variable sample based on the R data. The colour coding indicates the $M$ index and the symbol size the Stetson~J variability index, in line with the other figures. As discussed above, the most common variability is symmetric, around $M$~=~0. However, it can be seen that amongst the symmetric light curves, there are more objects with positive values for $M$ than with negative ones. There is a more or less homogeneous distribution of $Q$ values in the range from 0.0 to 0.6. A much larger fraction of sources has higher $Q$-values. Typically, sources with larger Stetson~J indices can have more asymmetric light curves. About 1.5~--~3.0 percent of the variables have $Q$ values outside the nominal range between zero and one, and are hence not shown. The reason for these is the underestimation of the photometric uncertainties towards fainter magnitudes, as discussed in Sect.\,\ref{lc_analysis}.

\subsubsection{Near and mid-infrared colours} \label{nir_col_all}

\begin{figure}
\centering
\includegraphics[angle=0,width=\columnwidth]{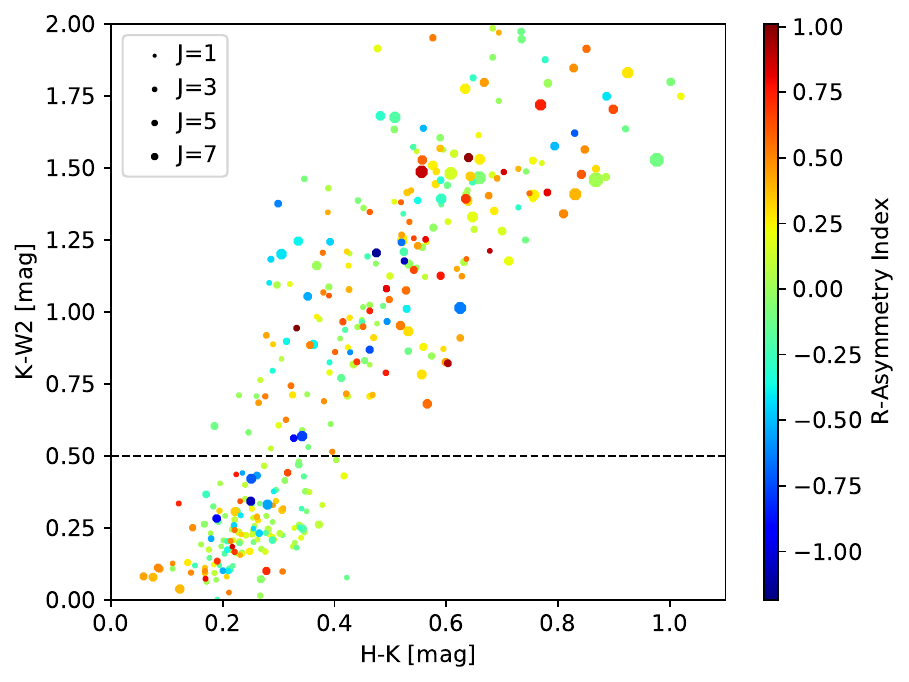} \hfill
\caption{ Near- and mid-infrared colour-colour plot using 2MASS and WISE data for all investigated variable cluster members. The dashed line marks the separation between objects with a detectable inner disk (above) and without (below). The colour code represents the R-band asymmetry index $M$ and the symbol size is proportional to the R-band Stetson~J index for each source. The legend gives an indication of the respective symbol sizes. \label{res_col}}
\end{figure}

In Fig.\,\ref{res_col} we show the distribution of near and mid infrared colours of the full sample of variable stars investigated, which have a detection at those wavelengths with a minimum SNR of 10. The colour code of the symbols indicates the asymmetry index $M$ and the symbol size represents the Stetson~J variability index. There is a clear separation of the objects into two groups. One group shows a clear K-W2 excess emission, i.e. represents sources with detectable inner disks. The other group has near zero mid-infrared colours, hence represents sources with no detectable inner disk, i.e. photospheres. The dashed line in Fig.~\ref{res_col} indicates the dividing line between these two groups at K-W2~$= 0.5$~mag \citep{2012A&A...540A..83T}. As discussed in the last section, both groups of sources are dominated by symmetric light curves. They also contain a mix of dippers and bursters, but there is an over abundance of dippers compared to bursters in general. 

On average our full sample contains 32 percent of objects with a K-W2 excess and 68 percent without. However, if only the variable sample of stars is considered (the ones shown in Fig.~\ref{res_col}), the fraction of sources with detectable inner disks increases to 60~--~65 percent, depending on the filter for which the Stetson~J index is determined. If we consider only objects that are variable in all three filters, then the disk fraction is 71 percent. Amongst the variables, the symmetric light curves dominate the population as discussed above, with 70~--~80 percent. However, the dippers outnumber the bursters in the K-W2 excess variables by 3:1 (in V and R) and 10:1 in I. For the variables without K-W2 excess emission, the bursters outnumber the dippers by 3:2 in V, and the populations are equal the R and I-Band. Some of the burster light curves could be caused by flares. We note that the HOYS survey is not designed to detect these reliably and refer back to our comments on the removal of photometry outliers and unusual light curves in Sect.\,\ref{lc_analysis}. 

\subsubsection{Light Curve Periodicity}

\begin{figure}
\centering
\includegraphics[angle=0,width=\columnwidth]{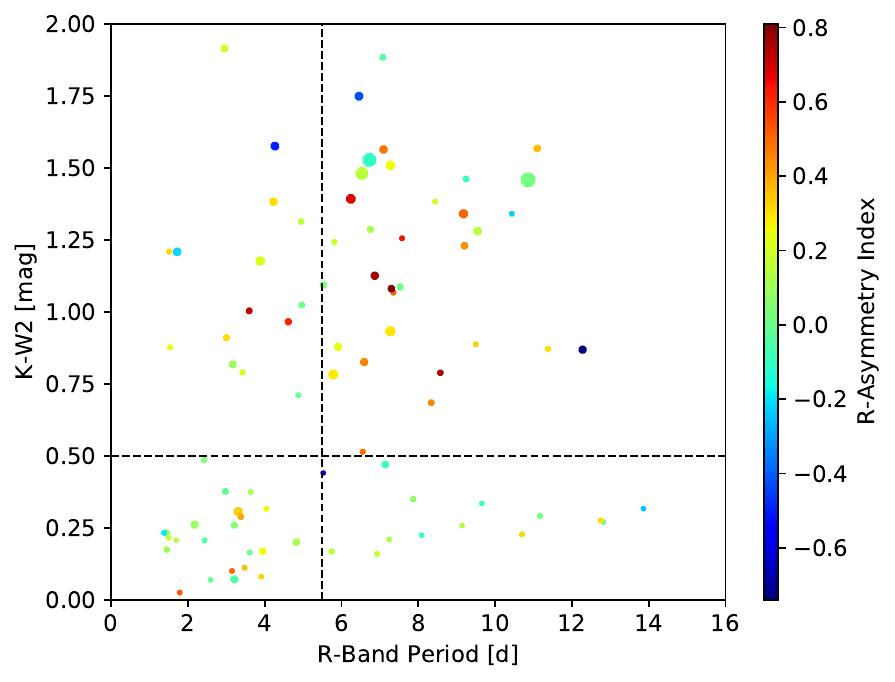} \\
\caption{ Significant periods in variable sources vs K-W2 colour. Sources above the horizontal dashed line are considered to have an excess in the K-W2 colour. The vertical dashed line separates fast from slow rotators. The symbol colour represents the $M$ index and the symbol size the Stetson~J variability index. \label{plot-per}}
\end{figure}

In Fig.\,\ref{plot-per} we show all variable sources for which a significant period has been identified. This period (determined from the R data) is plotted against the K-W2 colour. The colour coding and symbol size is the same as in the other figures. Note that the periods and associated FAPs have been determined for the entire, up to 7~yr long, light curves. Thus, spotted stars which show detectable periodic variations only during a part of their light curve, due to variable spot properties, are likely not included. Only sources with periodic variability, which is stable during the entire light curve are identified. We show the data (light curve, periodogram) for one such source in Fig.~\ref{phasefold} in the Appendix.

We find that amongst the periodic variables the fraction of K-W2 excess sources is 59/56/72 percent (for V, R, and I, respectively). This is roughly the same fraction that we found for all variable sources. Thus, the identification of periodic variability is not dependent on the presence of a detectable inner disk. Of all the variable stars, between 15 and 20 percent of the stars do have detectable periodic variability. Considering the above, these sources are hence most likely stars with stable, large, cold spots on their surface, or AA-Tau like objects with warped inner disks. The properties of the entire sample of periodic variables and their spots will be investigated in detail in Herbert et al. (in prep.). This will include a search for periodic signals in shorter (6 months) parts of each light curve.

There is a difference in how the fast and slow rotators are distributed amongst the variables with and without K-W2 excess. For the inner disk sources, between 65 and 70 percent (depending on the filter) are slow rotators, while 30 to 35 percent are fast rotators. Hence, the slow rotators outnumber the fast rotators by about 2:1 if the sources have a detectable inner disk. The situation is reversed for the objects without K-W2 excess. Here, between 57 and 81 percent are fast rotators, while between 19 and 43 percent have longer rotation periods. Thus, the fast rotators outnumber the slow rotators by 2:1. Compared to the investigation of IC~5070 by \citet{2021MNRAS.506.5989F}, these distributions match within the statistical uncertainties, but are based on a much larger sample. We note however, that the period identification for our current paper has been done using only one methodology and uses the entire 7~yr long light curves. The Herbert et al. (in prep.) paper will include a full discussion of this aspect. This change of the ratio of fast to slow rotators for objects with and without inner disks is consistent with the idea of disk breaking \citep[see e.g. the review by][]{2007prpl.conf..297H}. 

The periodic variables have preferentially (86 percent, same in all filters) symmetric light curves. This is expected if most periodic variations are due to spotted objects. Amongst the remaining sources, the dippers outnumber the bursters by about 3:1. Thus, only a fraction of 2~--~6 percent of objects are periodic bursters. The much larger fraction of periodic dippers (8~--~13 percent) is most likely dominated by AA-Tau type sources. This is supported by the fact that the periodic dippers are almost exclusively found amongst the K-W2 excess sources.

\subsection{Cluster population analysis}\label{res_cluster}

Here we discuss the light curve analysis results by cluster. Only the 17 clusters/groups with more than 20 stars with HOYS light curves are considered. Furthermore, we merged the stars in clusters/groups that have a spatial and proper motion overlap. This leaves a total of 12 clusters/groups of stars for our analysis. These are: IC~5070~a and b, NGC~1333, $\sigma$-Ori, $\lambda$-Ori, NGC~7129, M~42 (we merged M~42, L~1641~N, V~898~Ori, YY~Ori, and V~555~Ori), IC~348, IC~5146, NGC~2264~a and b, and IC~1396 (we merged IC~1396~A and IC~1396~N). Unless indicated otherwise in the name, only the stars from the main, most populous group in each target region are included in this analysis.

We aim to compare the light curve properties of the stars in each cluster to investigate how they potentially depend on the cluster properties. For this purpose we estimate the disk fraction of stars in each cluster. This has been determined as the fraction of all potential Gaia detected cluster members which have K-W2 excess emission due to circumstellar dust in the inner disk. We consider only stars with a detection in the 2MASS K-band and the WISE W2 filter, with uncertainties below 0.1~mag. This disk fraction can be considered an indicator for the typical age of the cluster. As the K-W2 excess emission only traces the inner disks, our disk fractions have to be considered a lower limit. All our clusters are closer than 1~kpc. Thus, we approximately detect stars down to the same mass in all clusters. Thus, any systematic biases caused by a potential mass dependence of the disk fraction, or by incompleteness for low mass stars, are very similar in all clusters. 

\subsubsection{Cluster Variability Fraction}

\begin{figure}
\centering
\includegraphics[angle=0,width=\columnwidth]{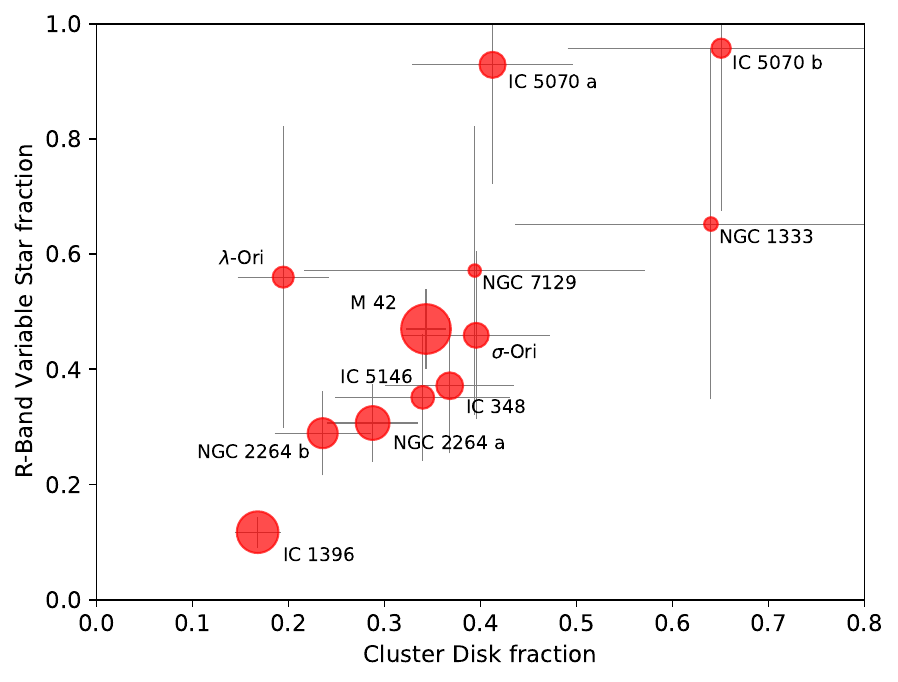} 
\caption{ Disk fraction vs. fraction of variable stars determined from the HOYS R-band light curves. The thin gray lines represent the uncertainties of the plotted ratios. The clusters are labeled as described in Sect.~\ref{res_cluster}. \label{var_frac}}
\end{figure}

We determine the fraction of stars that are variable in each cluster. This was done for each of the three filters (V, R, and I) separately. We show how this variability fraction compares to the disk fraction in each cluster in Fig.\ref{var_frac} for the R-band. The plots in the other filters do qualitatively look similar and some quantitative differences are discussed below. We have labelled the clusters with their common names and the circle size of the symbols is proportional to the log of the number of stars in the cluster used in our analysis. We over plot uncertainties for the determined fractions, which are based on the estimation that the uncertainty in the number ($N$) of sources is equal to $\sqrt{N}$.

In general (with some exceptions discussed below) the variability fraction correlates with the fraction of detectable inner disks in the cluster. Clusters with a higher inner disk fraction do show on average a larger fraction of variable stars. This indicates that younger clusters contain a larger fraction of variable stars. The correlation is stronger in the V and R filters, compared to the I-band. Furthermore, on average the variability fraction decreases with increasing wavelength of the observation. The latter is most likely caused by the fact that most causes for the photometric variability are stronger at shorter wavelengths. Both, accretion rate and line of sight extinction changes cause larger magnitude changes in V and R, compared to the I-band. Thus, stars are more likely to be detected as variable above the photometric uncertainty in those filters. This is in part counter-balanced by the fact that the young stars are fainter in the V and R filters, which in total leads to only a small dependence of the variability fraction on the filter used, especially for the most populous clusters. This correlation of disk and variability fraction is in line with the finding from Sect.\,\ref{nir_col_all} that the fraction of disk excess sources doubles amongst the variable stars compared to all cluster members in our sample.

There are some exceptions to the general trends seen in Fig.\,\ref{var_frac}. In NGC~1333 the variability fraction is slightly below the general trend. This could in part be a statistical effect since only 23 cluster members have a HOYS light curve with sufficient data points to analyse, and the uncertainties do support this. Both, the $\lambda$-Ori cluster and IC~5070~a are outliers above the general trend. They have a far higher variability fraction in V and R, with respect to the overall trend. Again, this could be a statistical effect, especially for $\lambda$-Ori as there are only 25 cluster members with a HOYS light curve in it, but not all of the deviation from the trend can be fully explained that way. One other possibility is that the correlation between variability fraction and disk fraction breaks down for very young samples with a disk fraction above 40\% (which includes two of the three outliers just mentioned).

\subsubsection{Other light curve properties}

We have investigated a number of other light curve properties for each sample of stars in the 12 clusters and groups. In particular these are the fractions of burster or dipper light curves amongst the variables sample, as well as the fraction of periodic or stochastic light curves. Typically there are some systematic trends of all these fractions depending on the filter used in obtaining the light curve. All of these are in line with the systematic biases in our data discussed above and the results obtained for the overall sample of stars analysed, including the occasional outliers from any trends.

For all the investigated light curve properties, there are however no correlations of any of the properties with the cluster disk fraction or any other cluster property. In other words, the dipper or burster fraction varies from cluster to cluster by a factor of a few, around the same average as found for the entire sample of variables, the values obtained for the same cluster in different filters correlate, but there are no correlations of those with any of the cluster properties. The same applies to the fraction of periodic or stochastic light curves. This might in part be due to the small number statistics in some clusters, but generally there seems to be indeed no significant trends even considering just the more populated clusters.

\section{Discussion and Conclusions}\label{discussion}

The HOYS project is providing long-term, optical, multi-filter, high-cadence monitoring of 25 target fields centred on presumed, nearby young star clusters and star forming regions. In this work we have utilised the Gaia~DR3 astrometry to identify potential cluster members and to analyse their optical light curves. A total of 45 groups/clusters with $\approx 17000$ Gaia selected members ($\approx 3000$ with well sampled HOYS light curves) have been identified. We selected 21 of these groups for detailed analysis, as they are within 1~kpc and clearly represent a population of young stars in the Gaia colour magnitude diagram. Thus, our detailed analysis includes 9143 potential cluster members, with well sampled optical light curves in V, R, and I for 1687 (18 percent) of them.

In many of the HOYS target fields, multiple populations of young stars were identified, and their distances and proper motions are summarised in Table~\ref{gaiatable}. For many of these this is the first time they are determined from Gaia~DR3 data. However, for several of the clusters our values are in agreement with previous works using earlier versions of Gaia data, such as DR2 and EDR3. These are for example: NGC~1333 and IC~348 \citep{2021MNRAS.503.3232P}; M~42 and other regions in the Orion complex \citep{2020A&A...643A.114C}; $\sigma$-Ori \citep{2022MNRAS.510.2883F}; IC~5070 \citep{2020ApJ...899..128K}; IC~1396 \citep{2023A&A...669A..22P}. Several of our regions are also included in \citet{2022A&A...664A..66M}.

Our work represents a large, homogeneous sample of nearby (within 1~kpc) young stars with long-term ($\approx 7$~yr) light curves in at least three optical filters (V, R, I). Similarly to the astrometric properties, in many cases this represents the first time such data have been made available and/or analysed. Previous works presenting the analysis of optical light curves (but mostly either using single filter data, or shorter survey duration) for a large number of members of some of our clusters include for example: NGC~1333 and IC~348 \citep[e.g.][]{2023RAA....23g5015W}; IC~348 \citep[e.g.][]{2000AJ....120..349H, 2004AJ....127.1602C, 2005AJ....130.2766K, 2006AJ....132.1555N, 2016MNRAS.462.2396F}; NGC~2264 \citep[e.g.][]{2014AJ....147...82C, 2016AJ....151...60S, 2022MNRAS.514.2736L}; $\sigma$-Ori \citep[e.g.][]{2004PhDT.........4S, 2004A&A...419..249S, 2010ApJS..191..389C}. Note that for many of the regions time series analysis has also been conducted in the infrared, such as for NGC~1333 \citep{2015AJ....150..175R}, IC~1396 \citep{2019ApJ...878....7M} and many others.

We identify light curves as variable if their Stetson~J index is above one. Depending on the filter used, the variability fraction varies. About one in four objects is considered variable in all three filters (V, R, and I), while almost half the objects are variable in at least one of the filters. The variability fraction in the sample increases with decreasing wavelength of the filter used. The asymmetry index $M$ of the light curves shows that roughly two thirds of all variable light curves can be classified as symmetric. The remainder are made up of dipper and burster light curves. For a sub-sample of bright sources (which is complete), the dippers outnumber the bursters by a factor of 2~$-$~4. Furthermore, the majority of light curves are classified as stochastic, based on the quasi-periodicity index $Q$. Strictly periodic light curves are rare and quasi-periodic light curves are uncommon.

The potential cluster members with reliable K and W2 photometry show a clear split into objects with and without detectable inner disks. They can be separated at K~$-$~W2~=~0.5~mag. Our full sample contains 32 percent  with disk excess emission and 68 percent without. Amongst the variable stars, however, the fraction of disk excess sources increases to 60~--~65 percent. In the disk excess sample, the dippers outnumber the bursters by just over three to one, while in the non disk excess sample the bursters and dippers occur in roughly the same amount. Note that the lack of disk excess emission only means that they are dust depleted inner disks. They can still be gas rich and have colder dust further out \citep[see e.g.][]{2014A&A...568A..18M}.

About one in six variables shows a detectable periodic variability. Sixty percent of these are disk excess objects. Thus, the detection rate of periodic variability is independent of the presence of a disk. The period distributions of objects with and without disks clearly support the idea of disk breaking \citep{2007prpl.conf..297H}. About two in three objects with detectable inner disk are slow rotators, while two thirds of the objects without disk are fast rotators.

As we found that the fraction of variable stars increases amongst the disk excess population, we investigated this on a cluster by cluster basis. We find that there is a clear correlation of the disk fraction in a cluster and the variability fraction. There are, however, a few outliers to this trend, which are explainable by small number statistics.
\section*{Acknowledgements}

We would like to thank all contributors of observational data for their efforts towards the success of the HOYS project.
This research has been partially supported by the supercomputing infrastructure of the NLHPC (ECM-02) in Chile. S. Vanaverbeke acknowledges the help of the IT support team of the NLHPC while working on the calculations described in this paper.
CH is supported by the Science and Technology Facilities Council under grant number STFC Kent 2020 DTP ST/V50676X/1.
JCW is funded by the European Union (ERC, WANDA, 101039452). Views and opinions expressed are however those of the author(s) only and do not necessarily reflect those of the European Union or the European Research Council Executive Agency. Neither the European Union nor the granting authority can be held responsible for them.
This work has made use of data from the European Space Agency (ESA) mission {\it Gaia} (\url{https://www.cosmos.esa.int/gaia}), processed by the {\it Gaia} Data Processing and Analysis Consortium (DPAC, \url{https://www.cosmos.esa.int/web/gaia/dpac/consortium}). Funding for the DPAC has been provided by national institutions, in particular the institutions participating in the {\it Gaia} Multilateral Agreement.

\section*{Data Availability Statement}

The data underlying this article are available in the HOYS database at http://astro.kent.ac.uk/HOYS-CAPS/.


\bibliographystyle{mnras}
\bibliography{bibliography} 


\appendix

\section{Notes on individual clusters in the manual Gaia selection}\label{app_clusters}

In this section we briefly discuss the clusters and groups of astrometrically coherent populations of stars identified in the HOYS target fields. Each section is identified by the name of the main target cluster. If uncertainties for distances or proper motions are quoted, then these are the standard errors for the mean as in Table\,\ref{gaiatable} and not the RMS scatter of the cluster members from the median. All comparisons to isochrones are based on the PHOENIX models by \citet{2013A&A...553A...6H}.

\subsection*{NGC~1333}

There is only one main peak in the histogram of Gaia distances and only one coherent cluster in proper motion space. Thus, this is a simple field. This main cluster has about the same proper motion ($\mu_\alpha$~=~$+$7.09~$\pm$~0.12~mas/yr; $\mu_\delta$~=~$-$9.92~$\pm$~0.07~mas/yr) and distance ($d$~=~298.5~$\pm$~1.7~pc) as the spatially distributed foreground group in the IC~348 field (see below). Hence, the same distributed population might exist in this field, but might not be identifiable due to this overlap and the small number of members. The scatter and uncertainty of $\mu_\alpha$ is about 70 percent larger than $\mu_\delta$. This might hint at a mix of populations in the cluster with slightly different proper motions. The same is the case for the foreground population in the IC~348 field (see below).

\subsection*{IC~348}

Similar to NGC~1333, there is only one single identifiable peak in the histogram of the Gaia distances. However, in the proper motion distribution there are two, well separated coherent groups. The more populated one, representing the main IC~348 cluster, has proper motions of $\mu_\alpha$~=~$+$4.48~$\pm$~0.04~mas/yr and $\mu_\delta$~=~$-$6.35~$\pm$~0.04~mas/yr. Its distance is $d$~=~318.1~$\pm$~1.2~pc. This places is about 20~pc further away than NGC~1333. The second coherent group is much less populated and clearly separated from the main cluster in proper motion space with $\mu_\alpha$~=~$+$6.72~$\pm$~0.10~mas/yr and $\mu_\delta$~=~$-$9.74~$\pm$~0.06~mas/yr. The distance of this group is $d$~=~297.0~$\pm$~4.0~pc. This places these stars at almost the same distance and proper motion as the NGC~1333 cluster. Furthermore, the members of the second group are homogeneously distributed across the field and thus represent a larger scale foreground population. Indeed this group is corresponding to the population identified and named Alcaeus in \citet{2021MNRAS.503.3232P}, using Gaia~DR2 data.

\subsection*{$\lambda$-Ori}

The distance histogram of this field only shows a single peak. However, in proper motion space there are two clear coherent populations. Their total proper motions differ by about 0.9~mas/yr. Most of that difference is in Right Ascension. The more populated, and hence main group of stars has $\mu_\alpha$~=~$+$0.78~$\pm$~0.02~mas/yr and $\mu_\delta$~=~$-$2.01~$\pm$~0.02~mas/yr, while the second group has $\mu_\alpha$~=~$+$1.65~$\pm$~0.02~mas/yr and $\mu_\delta$~=~$-$2.14~$\pm$~0.02~mas/yr. The distances of both groups are almost identical with $d$~=~398.9~$\pm$~1.6~pc for the main group and $d$~=~403.7~$\pm$~1.9~pc for the secondary group. The main difference between the groups is that the main one is clustered around $\lambda$-Ori itself, while the other objects are far more distributed, mostly to the South of the main group. 

\subsection*{M~42}

The cluster properties for M~42 (Orion Nebula) have been determined only from Gaia sources within 0.3~deg from the cluster centre. There is only a single peak in the distance histogram, and only one coherent group in proper motion space. While the mean proper motion ($\mu_\alpha$~=~$+$1.31~$\pm$~0.03~mas/yr, $\mu_\delta$~=~$+$0.22~$\pm$~0.04~mas/yr) can be evaluated quite accurately due to the large number of member stars, the scatter around that mean is quite large. The population of objects in proper motion space is also quite inhomogeneous with a lot of sub-structure, hinting at a number of overlapping sub-groups in the cluster. There is also a large number of objects with proper motions far away from the mean, potentially representing stars in the process of being ejected from the cluster. The median distance of the stars is $d$~=~400.4~$\pm$~0.9~pc. However, if one includes stars from the entire 0.6~deg field, then this distance drops by about 5~pc. This suggests that the more distributed populations just North (see V~555~Ori) and South (see L~1641~N), which overlap the M~42 field, are closer to Earth than the main cluster. We note that all the groups identified in all the HOYS fields covering the Orion star forming complex ($\lambda$-Ori, M~42, L~1641~N, $\sigma$-Ori, NGC~2068, V~898~Ori, YY~Ori, V~555~Ori), do coincide with stellar groups identified by \citet{2020A&A...643A.114C} based on the Gaia~DR2 data set. 

\subsection*{L~1641~N}

To avoid including any stars from the neighbouring YY~Ori field (see below), we only selected stars within 0.3~deg from the field centre for the determination of the distance and proper motions for this field. The distance of the stars in this field is $d$~=~388.4~$\pm$~1.3~pc, i.e about 12~pc closer than M~42 (see above). In the distribution of proper motions, we find a main compact group and two other less populated groups which are up to 1.5~mas/yr off-set. However, due to the small number of members in those, they do not show a clear sequence in a Gaia colour magnitude plot. We hence determine the median proper motion for all stars in those groups together ($\mu_\alpha$~=~$+$1.03~$\pm$~0.04~mas/yr, $\mu_\delta$~=~$+$0.33~$\pm$~0.05~mas/yr), but note that this region has potential sub-structure and should be investigated in more detail. The average proper motions in this field overlap the range found for the M~42 field, just to the North, and the YY~Ori field to the South..

\subsection*{$\sigma$-Ori}

This field has already been analysed in detail in \citet{2022MNRAS.510.2883F} using GaiaEDR3 data. Hence, our results are very similar to these. The histogram of distances shows two clear peaks, a main one containing the majority of stars and a secondary foreground population at a distance of $d$~=~368.5~$\pm$~1.7~pc. This foreground population shows a very compact distribution (RMS scatter of only 0.17~mas/yr) in proper motion space at $\mu_\alpha$~=~$+$1.80~$\pm$~0.02~mas/yr and $\mu_\delta$~=~$-$1.48~$\pm$~0.02~mas/yr. The sequence in the colour magnitude diagram indicates a slightly older population of stars. These objects are also widely distributed across the field, and hence represent a foreground moving group. 

The main peak in the distance histogram splits into two coherent groups in proper motion space. The main group is spatially clustered around $\sigma$-Ori at a distance of $d$~=~404.5~$\pm$~1.4~pc. The proper motion of this group is $\mu_\alpha$~=~$+$1.62~$\pm$~0.03~mas/yr and $\mu_\delta$~=~$-$0.49~$\pm$~0.03~mas/yr. There are hints of further sub-structure in the proper motion distribution of this group which should be looked at in more detail. The second group of stars in the main distance histogram peak has a proper motion of $\mu_\alpha$~=~$-$2.14~$\pm$~0.06~mas/yr and $\mu_\delta$~=~$+$1.27~$\pm$~0.07~mas/yr. Thus, it is clearly different from the main group and also spatially distributed. The median distance of the stars is $d$~=~421.9~$\pm$~3.7~pc. Thus, this might represent another group of young stars, about 20~pc in the background of the main $\sigma$-Ori cluster.

\subsection*{NGC~2068}

The field of NGC~2068 (or M~78) shows two peaks in the distance histogram. One peak at $\sim$250~pc and one at $\sim$410~pc. The stars in the 250~pc peak show no coherent structure in proper motion space, hence this is not a foreground group/cluster. The stars in the 410~pc peak split into three coherent groups in proper motion space. The two more populated groups (a and b) are close to each other with proper motions of $\mu_\alpha$~=~$-$0.98~$\pm$~0.05~mas/yr, $\mu_\delta$~=~$-$1.02~$\pm$~0.04~mas/yr and $\mu_\alpha$~=~$+$0.07~$\pm$~0.05~mas/yr, $\mu_\delta$~=~$-$0.56~$\pm$~0.04~mas/yr, respectively. Similarly, the distances are $d$~=~420.0~$\pm$~2.3~pc for group a and $d$~=~417.9~$\pm$~2.4~pc for group b. Spatially, group a  corresponds to the northern sub-cluster, while stars from group b are mostly located in the southern sub-cluster. The third group (c) is off-set in proper motion space with $\mu_\alpha$~=~$-$2.88~$\pm$~0.05~mas/yr and $\mu_\delta$~=~$+$1.19~$\pm$~0.07~mas/yr. These stars are spatially distributed and seem to align with a roughly 10~Myr isochrone in a colour magnitude diagram. At a median distance of $d$~=~379.3~$\pm$~6.9~pc they seem to form an older foreground moving group.

\subsection*{NGC~2244}

In this field (Rosette Nebula) there are two peaks in the distance histogram, one at $\sim$640~pc and one near 1500~pc. The latter is representing the main cluster in the field. The members have a median distance of $d$~=~1512.7~$\pm$~3.9~pc and a proper motion of $\mu_\alpha$~=~$-$1.75~$\pm$~0.01~mas/yr and $\mu_\delta$~=~$+$0.22~$\pm$~0.02~mas/yr, and are concentrated towards the centre of the cluster. The foreground population has a median distance of $d$~=~658.3~$\pm$~4.5~pc, with a proper motion of $\mu_\alpha$~=~$-$1.71~$\pm$~0.03~mas/yr and $\mu_\delta$~=~$-$4.76~$\pm$~0.03~mas/yr. It aligns with a roughly 20~Myr old isochrone in the colour magnitude diagram and is spatially distributed towards the South of the field. It thus represents an older foreground moving group.

\subsection*{NGC~2264}

In the NGC~2264 (Christmas Tree Cluster) field, there is only one peak in the distance histogram. In proper motion space this population in the peaks splits into two coherent groups which are about 1.2~mas/yr apart (mostly $\mu_\alpha$). Both groups are well populated. The slightly more populated one (a) represents the main cluster around NGC~2264, while the second group (b) is spatially clustered in the South of the field near the Cone Nebula. Both groups have similar distances with $d$~=~732.0~$\pm$~2.1~pc for group a and $d$~=~737.3~$\pm$~2.7~pc for group b, respectively. Their median proper motions are $\mu_\alpha$~=~$-$1.60~$\pm$~0.01~mas/yr, $\mu_\delta$~=~$-$3.64~$\pm$~0.01~mas/yr and $\mu_\alpha$~=~$-$2.43~$\pm$~0.02~mas/yr, $\mu_\delta$~=~$-$3.68~$\pm$~0.01~mas/yr, respectively.

\subsection*{V898~Ori}

This is the most Southern of our target fields in Orion. It has a single peak in the distance histogram. The proper motion distribution shows three potential sub-groups in close vicinity to each other, up to 1.6~mas/yr separation. But all them individually have too few members to analyse them separately. The distance of the group of $d$~=~402.8~$\pm$~3.9~pc places it at about the same distance as M~42. However, the proper motions of $\mu_\alpha$~=~$+$0.22~$\pm$~0.07~mas/yr and $\mu_\delta$~=~$-$0.31~$\pm$~0.06~mas/yr are different. There is no spatial clustering of members and a lack of objects in the North-Western part of the field.

\subsection*{YY~Ori}

This target field is situated between the M~42 and L~1641~N fields. Thus, to avoid overlaps, the group properties are determined for stars within 0.3~deg from the field centre only. There is a single peak in the distance histogram and a single compact group is evident in the proper motion distribution. Though, there are a number of stars scattered significantly away from the main proper motion group. These are most likely members of the M~42 population, which shows a similar proper motion distribution. The main compact group has a proper motion of $\mu_\alpha$~=~$+$1.29~$\pm$~0.01~mas/yr and $\mu_\delta$~=~$+$0.55~$\pm$~0.01~mas/yr, with a median distance of $d$~=~386.9~$\pm$~0.9~pc, which places it foreground to M~42 in this astrometrically complex field.

\subsection*{ASASSN-13DB}

There are two peaks in the distance histogram, one at $\sim$360~pc and one at $\sim$760~pc. The stars in the more distant peak show no coherent structure in the proper motion distribution. Thus they are most likely field stars. The stars in the nearby group are distributed as two coherent proper motion groups, separated by about 1.1~mas/yr. The main group (a) has a proper motion of $\mu_\alpha$~=~$+$1.31~$\pm$~0.03~mas/yr and $\mu_\delta$~=~$-$0.96~$\pm$~0.04~mas/yr, while the secondary group (b) has a proper motion of $\mu_\alpha$~=~$+$0.37~$\pm$~0.04~mas/yr and $\mu_\delta$~=~$-$1.24~$\pm$~0.06~mas/yr. The main group (a) has a distance of $d$~=~387.3~$\pm$~3.2~pc, while the second group (b) is closer at $d$~=~346.2~$\pm$~5.3~pc. Both groups do not show any spatial clustering. However, stars in group a are mostly found in the North-East of the field, while stars in group b are placed more central.

\subsection*{V555~Ori}

This field is just North of M~42. Thus, to avoid overlap, only stars within 0.3~deg from the field centre are included in the determination of the cluster properties. There is a single peak in the distance histogram. The proper motion distribution of the stars has a very similar shape to the M~42 field. It is very extended with potential sub-structure. The median proper motions of the potential members are $\mu_\alpha$~=~$+$1.23~$\pm$~0.03~mas/yr and $\mu_\delta$~=~$-$0.28~$\pm$~0.04~mas/yr, very similar to M~42. However, their median distance is $d$~=~393.4~$\pm$~0.9~pc, placing them slightly closer than M~42.

If one analyses the entire 0.6~deg field  around the central coordinates, there is a not well populated ($\sim$15 members), but coherent sub-group visible in proper motion space. These stars align well with a 10~Myr isochrone in a colour magnitude diagram. Their proper motions are $\mu_\alpha$~=~$-$1.74~$\pm$~0.05~mas/yr and $\mu_\delta$~=~$+$1.40~$\pm$~0.06~mas/yr, and their distance is $d$~=~419.4~$\pm$~3.8~pc. They are spatially distributed and hence seem to correspond to a slightly older, background moving group.

\subsection*{Gaia~17~bpi}

In this field there is no clear peak in the distance histogram. We know from \citet{2018ApJ...869..146H} that the approximate distance of the source is 1.27~kpc. When using a distance range around that value for the selection of Gaia sources, a coherent group of stars is identifiable in proper motion space. Though the group aligns with a young isochrone in a colour magnitude diagram, there is a large amount of scatter. This is most likely caused by strong and variable extinction, given that the majority of objects are situated in the South-Western part of the field, which coincides with a dark cloud. This group of stars has a proper motion of $\mu_\alpha$~=~$-$1.00~$\pm$~0.05~mas/yr and $\mu_\delta$~=~$-$5.65~$\pm$~0.06~mas/yr, and a distance of $d$~=~1300~$\pm$~6.3~pc.

\subsection*{Gaia~19~fct}

There is a wide peak around $\sim$1200~pc in the distance histogram. In proper motion space these stars split into three coherent groups. One of them is compact (group a) while the other two are more distributed and close to each other. We thus merge them together into group b. The main group (a) is centred in the South-East of the field and has a proper motion of $\mu_\alpha$~=~$+$0.25~$\pm$~0.01~mas/yr and $\mu_\delta$~=~$-$0.16~$\pm$~0.01~mas/yr. The distance of this group is $d$~=~1127~$\pm$~4.4~pc. The secondary group has a proper motion of $\mu_\alpha$~=~$-$3.57~$\pm$~0.04~mas/yr and $\mu_\delta$~=~$+$1.02~$\pm$~0.03~mas/yr, with a distance of $d$~=~1231~$\pm$~7.6~pc. However, it does not show a coherent alignment with an isochrone in a colour magnitude diagram, possibly due to variable line of sight extinction in this field. The distances are in agreement with the roughly 1~kpc quoted for young stars in this field in e.g. \citet{2016ApJ...827...96F} and \citet{2019ApJS..240...26S}.

\subsection*{Gaia~19~eyy}

There is no reason for any clusters to be found in this field, as it has been centred on the periodically erupting Be-star Gaia~19~eyy \citep{2023MNRAS.520.5413F}. Indeed there is no peak in the distance histogram for this field. However, a more systematic search identified three coherent populations of stars in proper motion space. The main group (a) seems to be a cluster in the North-East of the field near the coordinates RA~=~127.45~deg and DEC~=~-41.9~deg. The stars have a proper motion of $\mu_\alpha$~=~$-$1.77~$\pm$~0.01~mas/yr and $\mu_\delta$~=~$+$1.11~$\pm$~0.01~mas/yr, and a distance of $d$~=~1406~$\pm$~5.5~pc. In the colour magnitude diagram the cluster members align along the main sequence. The second group (b) forms a more spatially loose grouping in the South of the field near the coordinates RA~=~127.70~deg and DEC~=~-41.9~deg. Their proper motion is $\mu_\alpha$~=~$-$5.41~$\pm$~0.02~mas/yr and $\mu_\delta$~=~$+$4.96~$\pm$~0.02~mas/yr, with a distance of $d$~=~1274~$\pm$~5.7~pc. Finally, the third group (c) seems to represent a more distributed population of stars that aligns with a reddened main sequence in the colour magnitude diagram. It has a proper motion of $\mu_\alpha$~=~$-$4.87~$\pm$~0.01~mas/yr and $\mu_\delta$~=~$+$4.48~$\pm$~0.01~mas/yr, with a distance of $d$~=~2488~$\pm$~25~pc.

\subsection*{MWSC~3274}

There is a single peak in the distance histogram for this region. The stars in the peak also form a single coherent group in proper motion space with $\mu_\alpha$~=~$-$1.15~$\pm$~0.01~mas/yr and $\mu_\delta$~=~$-$1.59~$\pm$~0.01~mas/yr. The median distance of the group is $d$~=~884.5~$\pm$~3.5~pc. In the colour magnitude diagram the stars follow a sequence that can be interpreted as a reddened older cluster of stars. In \citet{2013A&A...558A..53K} the object is listed as a 4.5~Myr old cluster at a distance of 1580~pc, which is about twice our distance.

\subsection*{P~Cyg}

In this field there are two known compact clusters, Dolidze~41 and IC~4996. Dolidze~41 is situated at RA~=~304.65~deg and DEC~=~37.75~deg. It has an apparent radius of 0.15~deg and a distance of $d$~=~4.6~kpc. Thus, with the limits in our signal to noise requirement for the parallax values in our analysis, the stars of this cluster are too far way. The cluster IC~4996 is situated at RA~=~304.135~deg and DEC~=~37.65~deg. It has a distance of 1.9~kpc and an age of about 9~Myr. This cluster is easily identified and we determine its proper motion as $\mu_\alpha$~=~$-$2.60~$\pm$~0.01~mas/yr and $\mu_\delta$~=~$-$5.35~$\pm$~0.01~mas/yr. The median distance of the cluster stars is $d$~=~2159~$\pm$~8.8~pc. 

The main cluster associated with P~Cyg (MWSC~3301), should be at a distance of 1457~pc and have an age of 5~Myr \citep{2013A&A...558A..53K}. There is no indication in the Gaia~DR3 data of this cluster. In other words, selecting stars with distances within about 200~pc from the supposed distance, there are no coherent groups of stars in proper motion space. There is a peak in the distance histogram of the field at about 540~pc, but again there is no coherent group in proper motion space for these sources either.

\subsection*{Berkeley~86}

The proper motion distribution of the region shows three coherent groups of sources. Two of then (a and b) are very close together with proper motions of $\mu_\alpha$~=~$-$3.45~$\pm$~0.01~mas/yr and $\mu_\delta$~=~$-$5.48~$\pm$~0.01~mas/yr for group a and $\mu_\alpha$~=~$-$3.25~$\pm$~0.01~mas/yr and $\mu_\delta$~=~$-$6.01~$\pm$~0.01~mas/yr for group b. The median respective distance for these groups are $d$~=~1839~$\pm$~7.8~pc for group a and $d$~=~1853~$\pm$~7.9~pc for group b. Hence these are two spatially close clusters with very similar proper motions. Group a stars are distributed in the centre of the field, while the second group is mostly located in the South-West of the area. The third group (c) seems to be a cluster in the North-West of the field. Its stars have a distance of $d$~=~1636~$\pm$~13~pc and proper motions of $\mu_\alpha$~=~$-$0.70~$\pm$~0.01~mas/yr and $\mu_\delta$~=~$-$3.08~$\pm$~0.01~mas/yr. In the colour magnitude diagram, the sequence shows a lot of scatter, indicating potentially variable line of sight extinction towards the group members.

\subsection*{IC~5070}

The distance histogram shows a large drop in numbers for distances greater than 850~pc. This indicates a high extinction layer at the distance of the IC~5070 (or Pelican Nebula) region. The proper motion distribution shows two coherent groups with similar proper motions. The more populous group a is at $\mu_\alpha$~=~$-$1.33~$\pm$~0.03~mas/yr and $\mu_\delta$~=~$-$3.08~$\pm$~0.03~mas/yr, while the second group b has a proper motion of $\mu_\alpha$~=~$-$0.98~$\pm$~0.04~mas/yr and $\mu_\delta$~=~$-$4.15~$\pm$~0.04~mas/yr. The distances of the two groups are $d$~=~832.4~$\pm$~2.7~pc
and $d$~=~824.7~$\pm$~4.4~pc, for a and b, respectively. The more populous group of stars is mostly situated in the West of the field, while the secondary group is in the centre of the field. Our groups a and b correspond to the groups D and C (respectively) identified by \citet{2020ApJ...899..128K} based on Gaia~DR2 data. 

\subsection*{IC~1396A}

There is a single large peak in the distance histogram of IC~1396~A (Elephant Trunk Nebula, or Tr~37). Similarly there is one well populated, coherent group of sources in the proper motion distribution. There are hints of further sub-clusters in the area, but all of them do have a limited number of members in the area investigated, and all are spatially distributed. They might represent further sub-populations in this large \hii\ region. The main group has a proper motion of $\mu_\alpha$~=~$-$2.42~$\pm$~0.02~mas/yr and $\mu_\delta$~=~$-$4.72~$\pm$~0.02~mas/yr, and a distance of $d$~=~942.0~$\pm$~2.4~pc. The members of the group are clustered to the East of the centre of the field, around the ionising star of the \hii\ region, HD~206267. Our group of stars corresponds to the main group A of objects identified in \citet{2023A&A...669A..22P} based on Gaia~EDR3 data. This work also presents a more detailed analysis of the substructure in this field (covering also IC~1396~N, see below). The slight differences in the determined distance might be caused by us only analysing the central one degree field of this much larger \hii\ region.

\subsection*{IC~1396N}

This field spatially overlaps with the IC~1396~A region (see above). Hence, a full analysis of the entire \hii\ region is needed for a complete picture about the astrometric properties of the young stars in this region \citep[see e.g.][]{2023A&A...669A..22P}. There are two sub-groups in the field. The main one has a proper motion of $\mu_\alpha$~=~$-$2.02~$\pm$~0.03~mas/yr and $\mu_\delta$~=~$-$3.96~$\pm$~0.04~mas/yr, and a distance of $d$~=~955.0~$\pm$~3.1~pc. These are slightly different to the values for IC~1396~A, indicating some spatial variations within this large ($\sim$25~pc diameter) \hii\ region. 

We identify a second coherent group (b) of objects in this field. These stars have a proper motion of $\mu_\alpha$~=~$-$6.14~$\pm$~0.03~mas/yr and $\mu_\delta$~=~$-$5.83~$\pm$~0.03~mas/yr, and a distance of $d$~=~619.8~$\pm$~2.7~pc. Thus, this is a population in the foreground. The members are spatially distributed and their properties coincide with one of the potential sub-groups discussed in IC~1396~A (see above). The proper motion distribution of this field is very structured and complex and a detailed analysis is beyond the scope of this work.

\subsection*{NGC~7129}

This area contains the know old open cluster NGC~7142 in the South-West of the field, at the coordinates RA~=~326.00~deg and DEC~=~65.77~deg. This can be easily identified and we find a proper motion of $\mu_\alpha$~=~$-$2.68~$\pm$~0.01~mas/yr and $\mu_\delta$~=~$-$1.36~$\pm$~0.01~mas/yr for the cluster members, as well as a distance of $d$~=~2564~$\pm$~8.5~pc. The main cluster of young stars has a proper motion of $\mu_\alpha$~=~$-$1.73~$\pm$~0.03~mas/yr and $\mu_\delta$~=~$-$3.38~$\pm$~0.05~mas/yr and a distance of $d$~=~910.0~$\pm$~4.1~pc. There are no other coherent groups of stars in the field.

\subsection*{IC~5146}

The main cluster in the field is clearly identifiable as a coherent group in proper motion space at $\mu_\alpha$~=~$-$2.85~$\pm$~0.03~mas/yr and $\mu_\delta$~=~$-$2.70~$\pm$~0.04~mas/yr. The distance of the members is $d$~=~784.8~$\pm$~5.1~pc. There are hints of substructure in the proper motion distribution, but no clear indication of e.g. several sub-groups. A second group of stars can be identified in the proper motion distribution. This group is spatially distributed in the South of the field with proper motion values of $\mu_\alpha$~=~$+$1.28~$\pm$~0.02~mas/yr and $\mu_\delta$~=~$-$2.38~$\pm$~0.01~mas/yr. The median distance of the stars is $d$~=~1531~$\pm$~16~pc. The distribution of these objects in the colour magnitude plot is in agreement with a slightly reddened main sequence. Thus, this could potentially be a background moving group.

\subsection*{FSR~408}

There is one main peak in the distance histogram, which splits into two distinct groups (a and b) in proper motion space. The more populated group a has a proper motion of $\mu_\alpha$~=~$-$0.88~$\pm$~0.02~mas/yr and $\mu_\delta$~=~$-$2.29~$\pm$~0.01~mas/yr, and a distance $d$~=~860.2~$\pm$~1.8~pc. It represents the main cluster in the centre of the field and also a smaller cluster to the East of the main group of stars. The proper motion distribution is not symmetric, hinting at more than one sub-population. The second group (b) has a proper motion of $\mu_\alpha$~=~$-$2.33~$\pm$~0.02~mas/yr and $\mu_\delta$~=~$-$2.48~$\pm$~0.02~mas/yr, and a distance $d$~=~875.6~$\pm$~3.8~pc. Thus, these stars are slightly in the background of the main cluster and they are spatially distributed in the entire field. They are most likely part of the general population of sources in Cep~OB3.

\begin{figure*}
\section{Example light curves}
\centering
\includegraphics[angle=0,width=\columnwidth]{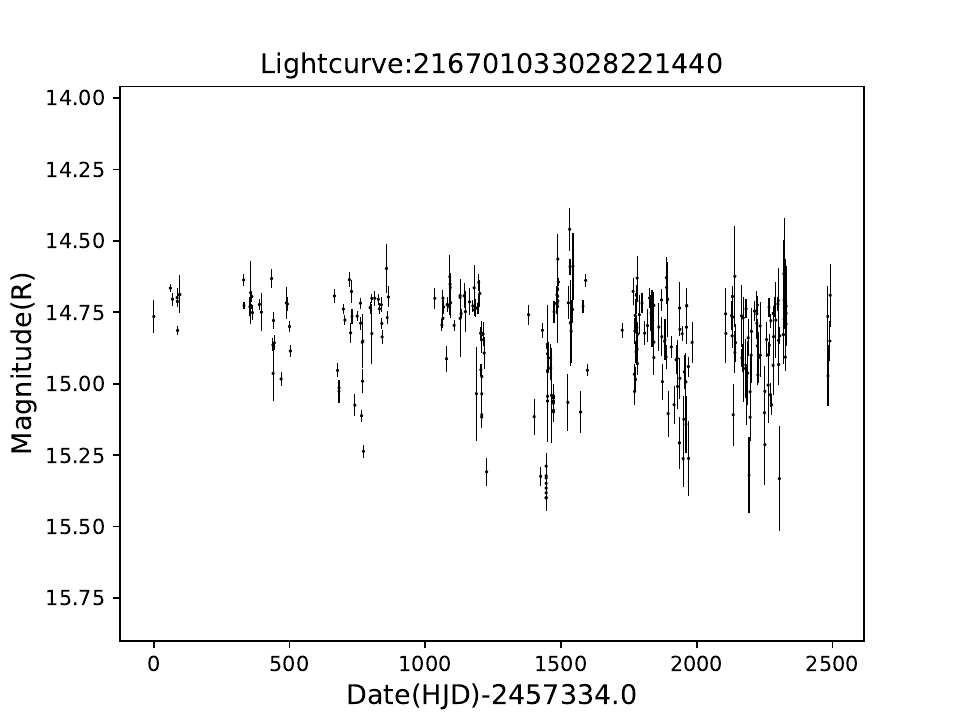} \hfill
\includegraphics[angle=0,width=\columnwidth]{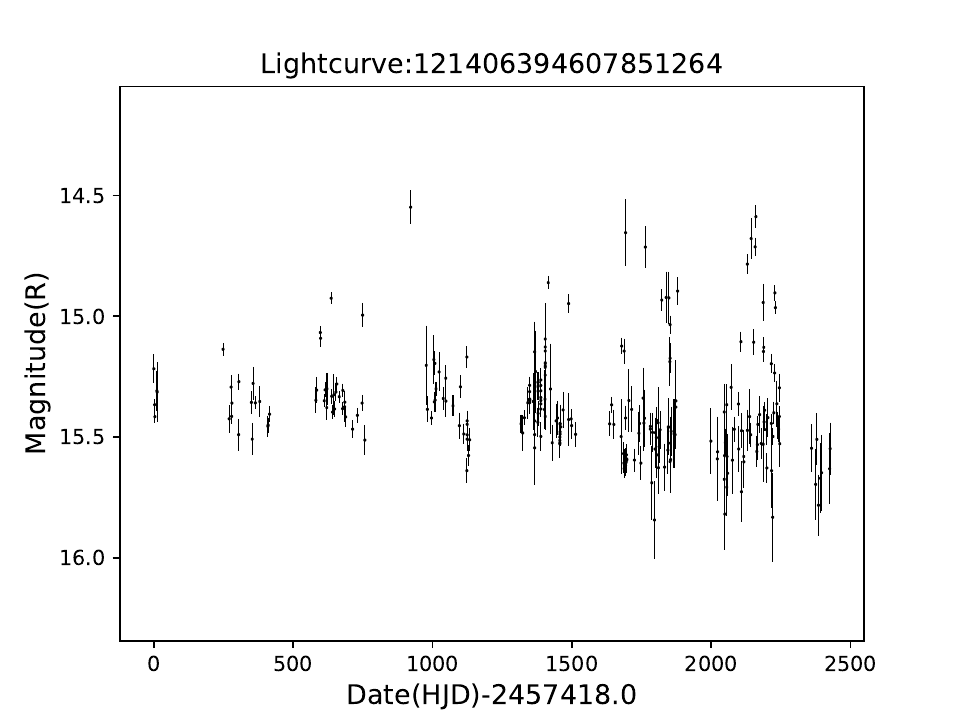} \\
\includegraphics[angle=0,width=\columnwidth]{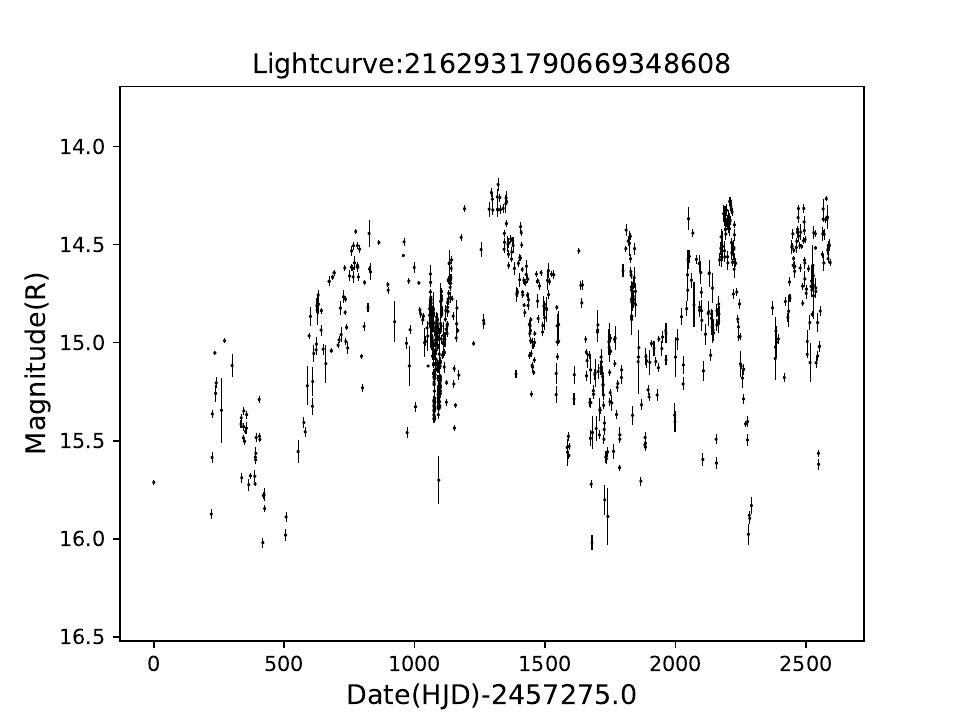} \hfill
\includegraphics[angle=0,width=\columnwidth]{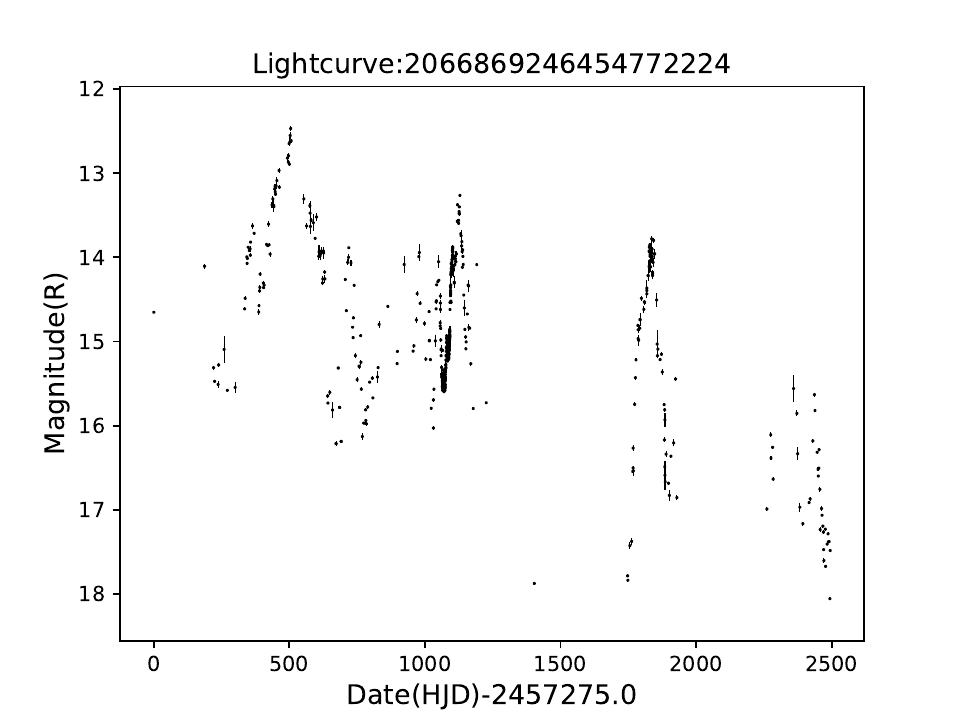} 
\caption{Example HOYS R-band light curves with photometric uncertainties for four non-periodic sources. {\bf Top Left:} Dipper, Gaia~DR3~ID~=~216701033028221440, J~=~2.127, M~=~0.736, Q~=~0.867; {\bf Top Right:} Burster, Gaia~DR3~ID~=~121406394607851264, J~=~2.672, M~=~$-$0.836, Q~=~0.909; {\bf Bottom Left:} Symmetric, Gaia~DR3~ID~=~2162931790669348608, J~=~11.569, M~=~0.159, Q~=~0.974; {\bf Bottom Right:} Symmetric, Gaia~DR3~ID~=~2066869246454772224, J~=~33.537, M~=~0.132, Q~=~0.975. \label{examplelc}}
\end{figure*}

\begin{figure*}
\section{Periodogram and Phase folded light curves}
\centering
\includegraphics[angle=0,width=\columnwidth]{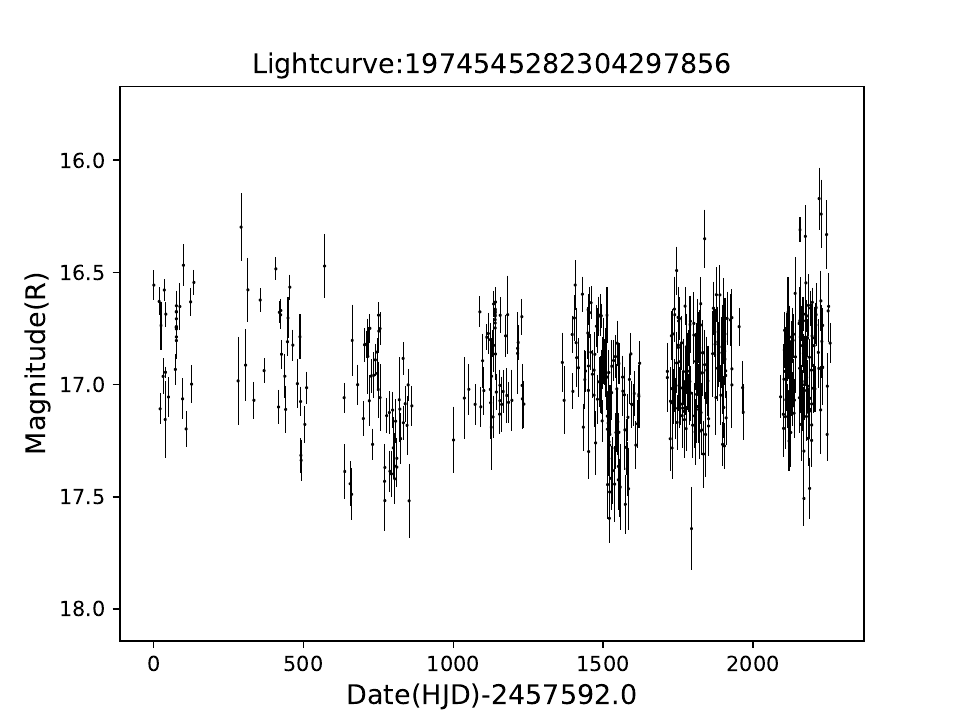} \hfill
\includegraphics[angle=0,width=\columnwidth]{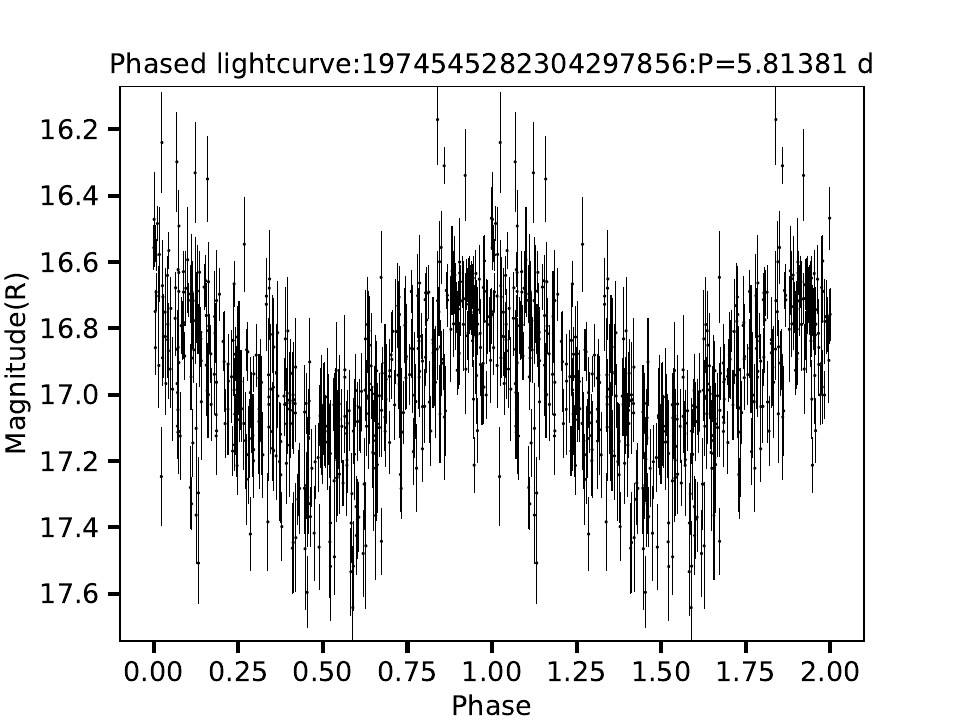} \\
\includegraphics[angle=0,width=\columnwidth]{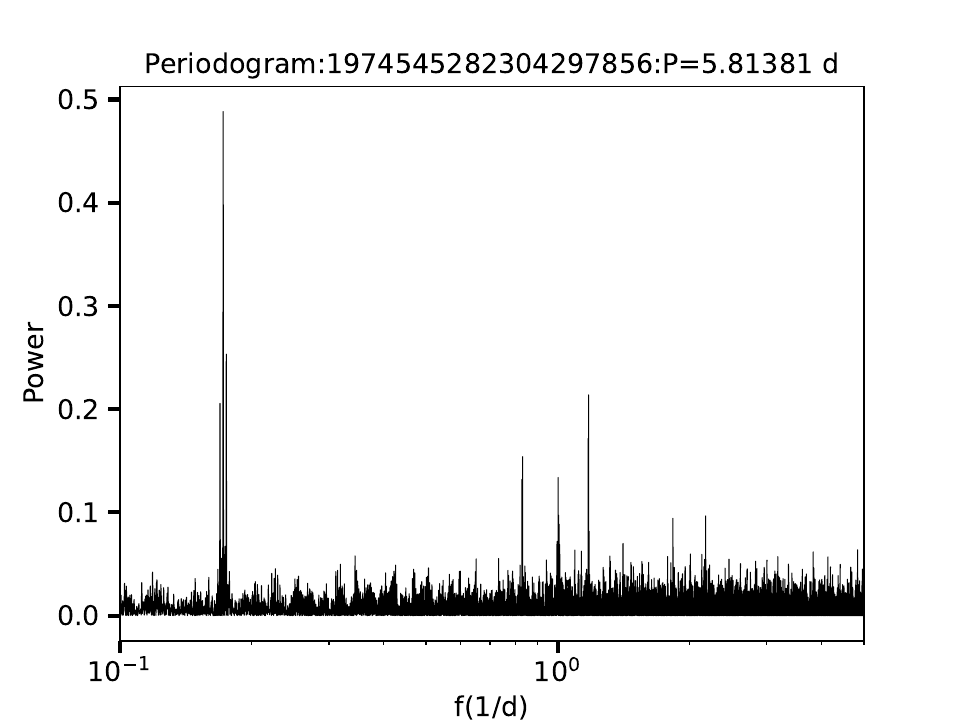} \hfill
\includegraphics[angle=0,width=\columnwidth]{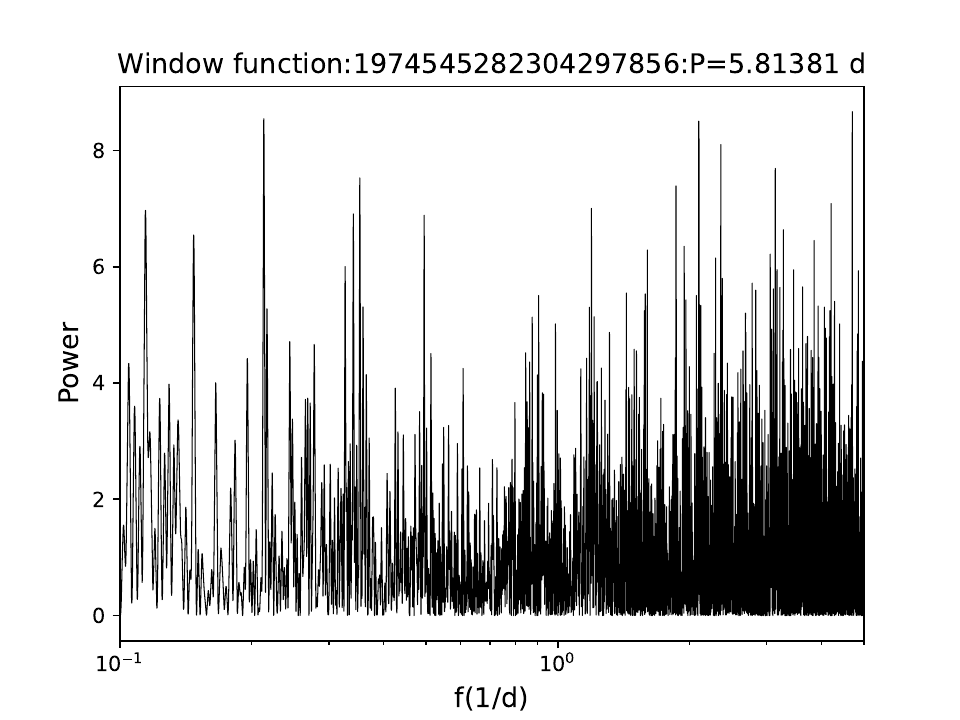} 
\caption{Example HOYS R-band data for a periodic source (Gaia~DR3~ID~=~1974545282304297856, Period~=~5.81381~d). The {\bf top left} panel shows the complete light curve. The {\bf top right} panel shows the phase folded light curve. The {\bf bottom left} panel shows the periodogram. The {\bf bottom right} panel shows the window function for this light curve. \label{phasefold}}
\end{figure*}

\bsp	
\label{lastpage}
\end{document}